\documentclass[a4paper,10pt]{article}

\usepackage{hyperref}
\usepackage[utf8]{inputenc}
\usepackage{amssymb}
\usepackage{amsmath}
\usepackage{graphicx}
\usepackage{xcolor}
\usepackage{natbib}
\usepackage{fancyhdr}

\usepackage{algorithm}
\usepackage{algpseudocode}
\usepackage{algorithmicx}

\usepackage[left=2cm,top=2.5cm,right=2cm,bottom=2.5cm]{geometry}

\hypersetup{colorlinks,linkcolor= {blue},citecolor= {blue}}

\graphicspath{{Figures/}}

\title{SMIwiz: An integrated toolbox for multidimensional seismic modelling and imaging}

\author{Pengliang Yang\\
  School of Mathematics, Harbin Institute of Technology, 150001, China\\
E-mail: ypl.2100@gmail.com}

\begin{document}

\maketitle
\begin{abstract}
    This paper contributes an open source software - SMIwiz, which integrates seismic modelling, reverse time migration (RTM), and full waveform inversion (FWI) into a unified computer implementation. SMIwiz has the machinery to do both 2D and 3D simulation in a consistent manner. The package features a number of computational recipes for efficient calculation of imaging condition and inversion gradient: a dynmaic evolving computing box to limit the simulation cube and a well-designed wavefield reconstruction strategy to reduce the memory consumption when dealing with 3D problems.
  The modelling in SMIwiz runs independently: each shot corresponds to one processor in a bijective manner to maximize the scalability. A batchwise job scheduling strategy is designed to handle large 3D imaging tasks on computer with limited number of cores.  The viability of SMIwiz is demonstrated by a number of applications on benchmark models.
\end{abstract}




{\bf PROGRAM SUMMARY}

\begin{small}
\noindent
{\em Program Title:} SMIwiz                                          \\
{\em CPC Library link to program files:} (to be added by Technical Editor) \\
{\em Developer's repository link:} \url{https://github.com/yangpl/SMIwiz} \\
{\em Code Ocean capsule:} (to be added by Technical Editor)\\
{\em Licensing provisions:} GNU General Public License v3.0 \\
{\em Programming language:} C, Shell, Fortran  \\
{\em Operating System:} Linux \\
{\em Software dependencies:} MPI, FFTW \\
{\em Nature of problem:}
Seismic modelling and imaging (FWI and RTM)\\
{\em Solution method:}
High-order finite-dfference time-domain (FDTD) for modelling on staggered grid; Quasi-Newton LBFGS algorithm for nonlinear optimization; line search to estimate step length based on Wolfe condition \\

\end{small}

\section{Introduction}

Seismic wave modelling and imaging are fundamental tools to understand the structure of the Earth in complex media.
Numerical simulation of wave propagation allows us to predict the response of the Earth to the given media properties. An efficient and memory frugal implementation of wave modelling is also the kernel of seismic imaging algorithms, such as reverse time migration (RTM) \citep{Baysal_1983_RTM,McMechan_1983_MET} and full waveform inversion (FWI) \citep{Tarantola_1984_ISR,Lailly_1983_SIP}.

The simulation of seismic waves can be done using different methods \citep{Virieux_2011_RSP}, i.e., finite difference method \citep{Virieux_1984_SWP,Virieux_1986_WPH,Levander_1988_FOF}, finite element method \citep{smith1975application}, pseudo-spectral method \citep{Carcione_2010_GFP}, spectral element method \citep{Seriani_1994_NSI,Komatitsch_1998_SEM} and discontinuous Garlerkin method \citep{hu1999analysis,dumbser2006arbitrary}. Among them, finite difference time-domain (FDTD) method gained its wide popularity due to small memory consumption, simple implementation and superior efficiency in a broad range of applications. The hardware acceleration such as graphics processing units (GPU) and OpenCL is particularly suitable for FDTD approaches to do numeric modelling \citep{abdelkhalek2009fast}.

Wave modelling and imaging are computationally intensive tasks. There are many open source codes implementing them in a number of independent ways. 
Great effort has been spent on efficient implementation of wave simulation. A parallel 3D viscoelastic seismic modelling using FDTD was conducted by  \cite{Bohlen_2002_PVF}. In SEISMIC\_CPML, a collection of Fortran90 programs were developed to solve 2D and/or 3D isotropic or anisotropic elastic, viscoelastic or poroelastic wave equation using FDTD combined with convolutional perfectly matched layer (PML) conditions \cite{Komatitsch_2007_GEO}. SOFI2D and SOFI3D \citep{bohlen2015sofi3d} are 2D and 3D seismic modelling coded in  C programming language. 3D elastic wave equations are also solved via FDTD method on parallel GPU devices \citep{Michea_2010_AFP,weiss2013solving}.

Using these modelling engines, two types of imaging are routinely conducted in exploration geophysics: one is the migration of seismic reflections to form an image of the subsurface, the other is building the velocity model according to observed seismograms. RTM has replaced a number of one-way migration techniques and becomes the standard technology for structural imaging \citep{Etgen_2009_ODI}. RTM assumes a known velocity with sufficient accuracy, which may be acquired using FWI-based velocity model building \citep{Virieux_2009_OFW}.

Inspired by the good convergence of quasi-Newton optimization,  TOY2DAC code was released for 2D frequency domain FWI \cite{Brossier_2011_TDF}. To study the influence of model parametrization,  2D time-domain isotropic (visco)elastic finite-difference modeling and FWI code DENIS has been developed for P/SV-waves  \cite{Kohn_2012_IMP}. Time-domain RTM and FWI are implemented on GPU devices with an effective boundary saving strategy \citep{Yang_2014_RTM,Yang_2015_GPU}. Unfortunately, these implementations were limited in 2D while the potential of MPI parallelization over distributed memory has not yet been fully exploited.
The SeisCL software utilizes OpenCL  to do time-domain seismic modeling in viscoelastic media for FWI \cite{fabien2017time}. The modelling was combined with MPI for domain decomposition. The test of SeisCL over different hardware architectures demands device specific optimization for good performance. Elastic RTM and FWI, despite their theoretic importance, continue to be academic exercises, and remains to be explored for commercial usage.

In this work, we focus on developing a FDTD-based open source software \verb|SMIwiz| for multidimensional seismic modelling and inversion. Based on acoustic wave equation, the modelling engine is coded in C, running over a number of parallel processors using distributed memory thanks to MPI-based high performance computing architecture. Each modelling job for independent sources is parallelized with OpenMP for speedup using local shared memory. \verb|SMIwiz| builds a machinery to do both 2D and 3D simulation in a consistent manner. The package features a number of computational recipes for efficient calculation of imaging condition and inversion gradient: a dynamic evolving computing box to limit the simulation cube, a well-designed wavefield reconstruction strategy to reduce the memory consumption typically in 3D, and a batchwise job scheduling strategy giving flexibility to run over heterogeneous computing architectures.

\section{Methodology}

\subsection{Forward modelling}

The first order isotropic acoustic wave equation is given by
\begin{equation}\label{eq:acoustic}
  \begin{cases}
    \rho(\mathbf{x}) \partial_t \mathbf{v}(\mathbf{x},t) + \nabla p(\mathbf{x},t) = 0,    &\mathbf{x}\in X, t\in [0,T]\\
    \partial_t p(\mathbf{x},t) + \kappa(\mathbf{x}) \nabla\cdot \mathbf{v}(\mathbf{x},t) = f_p(\mathbf{x},t), & \mathbf{x}\in X, t\in [0,T]
  \end{cases}
\end{equation}
where $\mathbf{v}=[v_x, v_y, v_z]^T$ is the particle velocity including three directional components $v_x$, $v_y$ and $v_z$; $p$ denotes the acoustic pressure; $f_p$ stands for an explosive source term; $\rho$ is the density; $\kappa$ is the bulk modulus which can be expressed via density $\rho$ and wave speed $V_p$, i.e., $\kappa=\rho V_p^2$. Clearly, the pressure and particle velocity fields are functions of both time $t$ and space $\mathbf{x}$, while the medium parameters $\rho$ and $\kappa$ vary only in space. 

\begin{figure}
  \centering
  \includegraphics[width=0.5\linewidth]{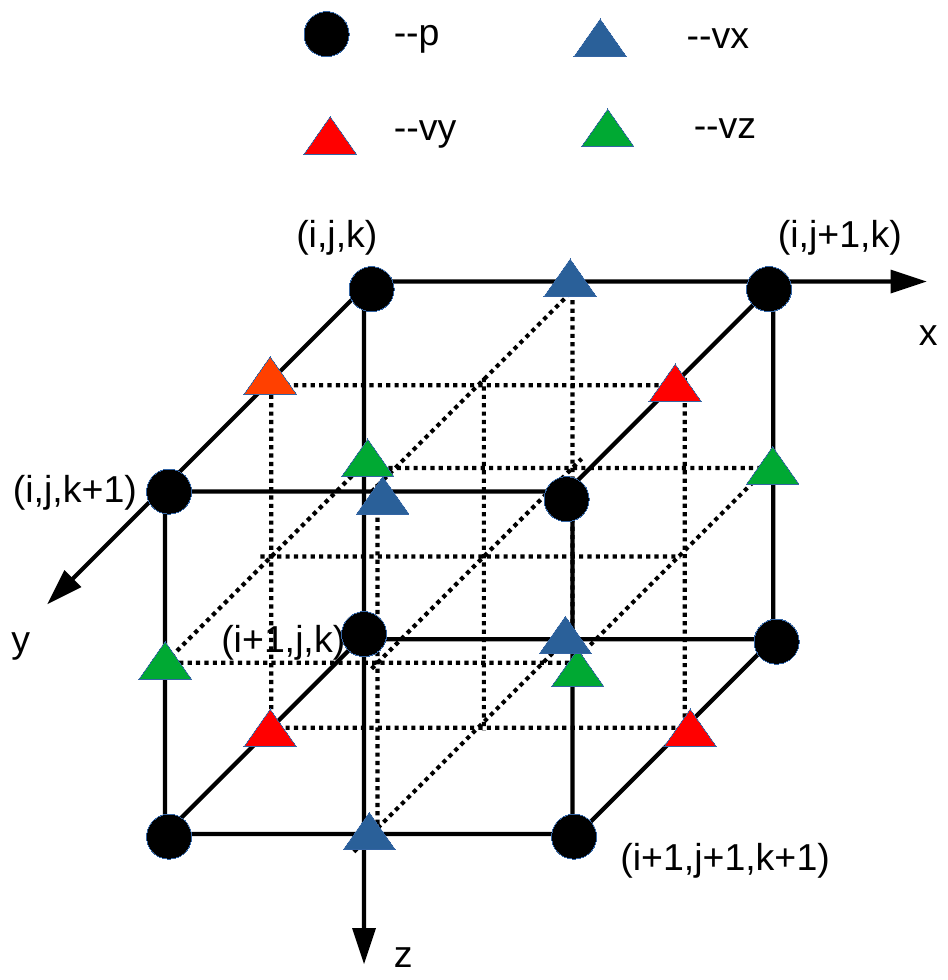}
  \caption{Pressure and particle velocities on the staggered grid}\label{fig:fdgrid}
\end{figure}

To do numerical simulation, we consider the simplest leap-frog finite difference method on staggered grid \citep{Virieux_1986_WPH,Levander_1988_FOF,Graves_1996_SSW}. Assume the time are discretized as $t_n=n\Delta t$, while the space variables are also discretized using equal grid spacing, i.e., $x_i=i\Delta x$, $y_j=j\Delta y$ and $z_k=k\Delta z$. As depicted in Figure~\ref{fig:fdgrid}, the pressure sits at integer grid at integer time step, that is, $p(x_i,y_j,z_k,t_n)= p(i\Delta x,j\Delta y, k\Delta z, n\Delta t)$; the particle velocities are defined at half integer time step and shifted half grid along x, y and z directions, i.e., $v_x(x_{i+1/2},y_j,z_k,t_{n+1/2})$, $v_y(x_i,y_{j+1/2},z_k,t_{n+1/2})$ and $v_z(x_i,y_j,z_{k+1/2},t_{n+1/2})$. For brevity, we denote them as follows
\begin{equation}
  \begin{cases}
    p(x_i,y_j,z_k,t_n) &\equiv p|_{i,j,k}^n\\
    v_x(x_{i+1/2},y_j,z_k,t_{n+1/2})&\equiv v_x|_{i+1/2,j,k}^{n+1/2}\\
    v_y(x_i,y_{j+1/2},z_k,t_{n+1/2})&\equiv v_y|_{i,j+1/2,k}^{n+1/2}\\
    v_z(x_i,y_j,z_{k+1/2},t_{n+1/2})&\equiv v_z|_{i,j,k+1/2}^{n+1/2}\\
  \end{cases}
\end{equation}

Based on equation \eqref{eq:acoustic} and the homogeneous initial condition, the fields can be advanced from one time step to the next:
\begin{equation}
  \begin{cases}
    v_x|_{i+1/2,j,k}^{n+1/2} &= v_x|_{i+1/2,j,k}^{n-1/2} -\Delta t \rho_{i+1/2,j,k}^{-1}\partial_x p|_{i+1/2,j,k}^{n},\\
    v_y|_{i,j+1/2,k}^{n+1/2} &= v_y|_{i,j+1/2,k}^{n-1/2} -\Delta t \rho_{i,j+1/2,k}^{-1}\partial_x p|_{i,j+1/2,k}^{n},\\
    v_z|_{i,j,k+1/2}^{n+1/2} &= v_z|_{i,j,k+1/2}^{n-1/2} -\Delta t \rho_{i,j,k+1/2}^{-1}\partial_x p|_{i,j,k+1/2}^{n},\\
    p|_{i,j,k}^{n+1} &= p|_{i,j,k}^n -\Delta t \kappa_{i,j,k} (\partial_x v_x + \partial_y v_y + \partial_z v_z)|_{i,j,k}^{n+1/2} + \Delta t f_p|_{i,j,k}^{n+1/2},
  \end{cases}
\end{equation}
where the first spatial derivatives are approximated using high order finite difference schemes over staggered grid. For x direction, the first derivatives approximated by finite difference of order $O(2N)$  are
\begin{equation}
  \begin{cases}
    \partial_x v_x|_{i,j,k} = \sum_{l=1}^N a_l (v_x|_{i+(2l-1)/2,j,k} - v_x|_{i-(2l-1)/2,j,k})/\Delta x\\
    \partial_x p|_{i+1/2,j,k} = \sum_{l=1}^N a_l (p|_{i+l,j,k} - p|_{i-l+1,j,k})/\Delta x\\
  \end{cases}
\end{equation}
where $a_l$, $l=1,\cdots, N$ are FD coefficients which can be found in \cite{Fornberg_1998_NFD}. Based on von Neumann analysis, the following Courant-Friedrichs-Lewy (CFL) stability condition must be satisfied
\begin{equation}\label{eq:cfl}
  \Delta t V_{p\max} \sqrt{\frac{1}{\Delta x^2} +\frac{1}{\Delta y^2} +\frac{1}{\Delta z^2} }\sum_{l=1}^N |a_l|\leq 1.
\end{equation}

To mimic the waves propagating to infinity, we add 20 perfectly matched layers (PMLs) surrounding the domain of interest \citep{Komatitsch_2007_GEO}. This avoids the artificial reflections due to  the truncation of the computing mesh. On the top of the boundary, we have implemented free surface boundary using the image method, by asymmeric mirroring of the pressure and symmetric mirroring of the vertical component of the particle velocity  according to \cite{Levander_1988_FOF}. The injection of the seismic source and extraction of the modelled data at arbitrary location is performed by Kaiser-windowed sinc interpolation \citep{Hicks_2002_ASR}.

\subsection{Nonlinear full waveform inversion}

Full waveform inversion aims to iteratively reconstruct the true subsurface parameters by minimizing the error between the observed seismogram $d$ and the synthetic data $u$ predicted according to the wave equation and the given parameters. The misfit functional for FWI is defined in least-squares sense
\begin{equation}\label{eq:objfwi}
  J(m) = \frac{1}{2} \sum_s\sum_r \int_0^T (u(\mathbf{x}_r,t;\mathbf{x}_s)-d(\mathbf{x}_r,t;\mathbf{x}_s))^2 \mathrm{d}t,
\end{equation}
where the inversion parameter $m$ is a function of $\rho$ and $\kappa$; $u(\mathbf{x}_r,t;\mathbf{x}_s)$ and $d(\mathbf{x}_r,t;\mathbf{x}_s)$ are synthetic and observed seismograms recorded at receiver $\mathbf{x}_r$ associated with source $\mathbf{x}_s$.  The gradient of the misfit with respect to density and bulk modulus are 
\begin{equation}\label{eq:gradrhokappa}
  \begin{cases}
    \frac{\partial J(m)}{\partial \rho}=\int_0^T \bar{\mathbf{v}}^\mathrm{H}\partial_t \mathbf{v} \mathrm{d}t\\
    \frac{\partial J(m)}{\partial \kappa}=\frac{1}{\kappa}\int_0^T \bar{p}\nabla\cdot \mathbf{v}\mathrm{d}t
  \end{cases},
\end{equation}
where $(\bar{\mathbf{v}},\bar{p})$ are the adjoint fields satisfying
\begin{equation}\label{eq:adj}
  \begin{cases}
    \rho\partial_t \bar{\mathbf{v}} + \nabla \bar{p} = -\delta d_\mathbf{v},\\
    \partial_t\bar{p} + \kappa\nabla\cdot \bar{\mathbf{v}} = -\kappa \delta d_p,
  \end{cases}
\end{equation}
in which $\delta d_\mathbf{v}$ and $\delta d_p$ are data residuals corresponding to particle velocity and pressure components. Since the adjoint equation \eqref{eq:adj} shares exactly the same structure as the forward state equation \eqref{eq:acoustic} except the source terms, we can use the same solver to model $(\bar{\mathbf{v}}, \bar{p})$. The inversion may be carried out using different model parametrizations, for example, $(V_p,\rho)$ or $(V_p,I_p)$ where the P-wave velocity $V_p$ and impedance $I_p$ are related to density via $I_p=\rho V_p$. The misfit gradient with respect to other parameters can easily be deduced from equation \eqref{eq:gradrhokappa} thanks to the chain rule. We refer to \cite{Yang_2024_formulation} to check out the derivation of the specific expressions and the mathematical details. In order to minimize the misfit $J(m)$, FWI iterates over the model parameter via
\begin{equation}
  m^{k} = m^{k-1} + \alpha \delta m,
\end{equation}
where $\alpha$ is a proper step length  estimated through line search method  \citep{Nocedal_2006_NOO} at the $k$-th iteration; $\delta m$ is the descent direction determined based on the gradient of the misfit functional $\nabla J(m)$. To achieve good convergence in nonlinear optimization, $\delta m$ can be computed through limited-memory Broyden–Fletcher–Goldfarb–Shanno (LBFGS) algorithm  or solving the normal equation using Newton-CG method \citep{Yang_2018_TRN}. In practice, \verb|SMIwiz| utilizes the two-loop recursion LBFGS implementation to find a descent direction $\delta m$ \citep{Nocedal_1980_UQN}. A pseudo code to implement the nonlinear FWI for model update is shown in Algorithm \ref{alg:lbfgs}.

\begin{algorithm}
  \caption{Nonlinear FWI for updating model parameters $m$}\label{alg:lbfgs}
  \begin{algorithmic}[1]
    \State Given initial model: $m_0$\;
    \For{$k=1,\cdots,niter$}
    \State calculate misfit gradient: $\nabla J$\;
    \State estimate approximate Hessian $\tilde{H}$ using LBFGS algorithm\;
    \State estimate descent direction: $\delta m=- \tilde{H}^{-1} \nabla J$
    \State find a good step size $\alpha$ by line search\;
    \State update model $m_{k}:=m_{k-1}+\alpha \delta m$\;
    \EndFor
  \end{algorithmic}
\end{algorithm}

\subsection{Reverse time migration}

Given a sufficiently accurate background velocity model, one can migrate the seismic events back to its true location where the scattering happens. A defacto technique to do it is the reverse time migration (RTM), which dates back to \cite{Baysal_1983_RTM,McMechan_1983_MET}  four decades ago. With the advance of computer power, RTM has become a mature and standard imaging technique to produce a structural image of the subsurface.  RTM overcomes the limitation of one-way imaging methods in handling the turning waves, high dip and lateral velocity contrast \citep{Zhang_2008_RTM}. It has a wide range of applications in both exploration geophysics \citep{Zhang_2009_PIR} and global seismology \citep{zhang2023nature}. The RTM image is simply the cross-correlation of the source field $p_s(\mathbf{x}, t;\mathbf{x}_s)$ (computed forward in time with a known source) and the receiver field $p_r(\mathbf{x}, t;\mathbf{x}_s)$ (simulated by simultaneously backpropagating the reflection data from multiple receivers)
\begin{equation}\label{eq:rtmimage}
  I(\mathbf{x}) = \sum_s \int_0^T p_r(\mathbf{x}, t;\mathbf{x}_s) p_s(\mathbf{x}, t;\mathbf{x}_s)\mathrm{d}t.
\end{equation}
To enhance the structural image at depth, one often consider the normalized imaging condition \citep{Yang_2014_RTM}, in which the images of each source are normalized by source energy/illumination. These imaging conditions easily suffer from low-frequency artifacts (the imprint of the wave path that the total traveltime of different waves coincides). The application of a Laplace filtering helps to eliminate these artifacts \citep{Zhang_2009_PIR}.

Note that the computation of imaging condition in equation \eqref{eq:rtmimage} follows exactly the same procedure as FWI gradient building, except the injection of the adjoint source is the observed data instead of data residual. Algorithm \ref{alg:fg} summarizes the steps for the evaluation of FWI function value and gradient (as well as RTM image).  Algorithm \ref{alg:fg} includes a number of useful tricks to achieve good efficiency. We will detail them in Sections~\ref{sec:software} and \ref{sec:implementation}.

\begin{algorithm}
  \caption{Evaluation of FWI function and gradient/RTM image}\label{alg:fg}
  \begin{algorithmic}[1]
    \State \verb|check_cfl(sim);|
    \State \verb|cpml_init(sim);|
    \State \verb|extend_model_init(sim);|
    \State \verb|fdtd_init(sim, 1);//flag=1, incident field allocation|
    \State \verb|fdtd_init(sim, 2);//flag=2, adjoint field allocation|
    \State \verb|fdtd_null(sim, 1);//flag=1, incident field initialization|
    \State \verb|fdtd_null(sim, 2);//flag=2, adjoint field initialization|
    \State \verb|decimate_interp_init(sim);|
    \State \verb|extend_model(sim);|
    \State \verb|computing_box_init(acq, sim, 0);|
    \State \verb|computing_box_init(acq, sim, 1);|
    \For{$it=0,\cdots,nt-1$(forward modelling)}
    \State \verb|decimate_interp_bndr(sim, 0, it);//interp=0|
    \State \verb|fdtd_update_v(sim, 1, it, 0);//update fwd v|
    \State \verb|fdtd_update_p(sim, 1, it, 0);//update fwd p|
    \State \verb|inject_source(sim, acq, sim->p1, sim->stf[it]);|
    \State \verb|extract_wavefield(sim, acq, sim->p1, sim->dcal, it);|
    \EndFor
    \State \verb|compute adjoint source (and FWI misfit - MPI_Allreduce)|

    \For{$it=nt-1,\cdots,0$(adjoint modelling)}
    \State \verb|inject_adjoint_source(sim, acq, sim->p2, sim->dres, it);|
    \State \verb|fdtd_update_v(sim, 2, it, 1);//update adj v|
    \State \verb|fdtd_update_p(sim, 2, it, 1);//update adj p|
    \State \verb|inject_source(sim, acq, sim->p1, sim->stf[it]);|
    \State \verb|fdtd_update_p(sim, 1, it, 0);//reconstruct fwd p|
    \State \verb|fdtd_update_v(sim, 1, it, 0);//reconstruct fwd v|
    \State \verb|decimate_interp_bndr(sim, 1, it);//interp=1|
    \State \verb|integrate FWI gradient/RTM image|
    \EndFor
    \State \verb|cpml_close(sim);|
    \State \verb|extend_model_close(sim);|
    \State \verb|fdtd_close(sim, 1);|
    \State \verb|fdtd_close(sim, 2);|
    \State \verb|decimate_interp_close(sim);|
    \State \verb|computing_box_close(sim, 0);|
    \State \verb|computing_box_close(sim, 1);|
    \State \verb|merge all local FWI gradient/RTM image - MPI_Allreduce|
  \end{algorithmic}
\end{algorithm}

\section{Software description}\label{sec:software}

\subsection{Generic design}

\verb|SMIwiz| performs 2D and 3D seismic modelling and imaging which runs over single CPU processor as well as multiple CPU processors in parallel. It requires 
\begin{itemize}
\item FFTW for fast Fourier transform;
\item MPICH or OpenMPI for multi-processor parallelization over distributed memory architecture.
\end{itemize}
It defines a number of pointers to ensure a clean and consistent implementation:
\begin{itemize}
\item The pointer of type \verb|acq_t| are specific to acquisition geometry. It reserves the number of source and receiver for each process as well as their coordinates in x-, y- and z- directions.
\item The pointer of type \verb|sim_t| is the simulator for forward and adjoint modelling. There are huge number of parameters inside, which will be detailed shortly.
\item The pointer of type \verb|fwi_t| is the inversion/imaging data structure, maintaining the number of unknown parameters and their bounds, the choice of parametrization and preconditioner, misfit value etc.
\item The pointer of type \verb|opt_t| is dedicated to nonlinear optimization. The relevant parameters for LBFGS, such as the number of iterations, the memory length of LBFGS, the step length of line search, as well as the vectors for storing the inversion parameters and gradients. In the framework of \verb|SMIwiz|, we have re-implemented LBFGS and Newton-CG algorithms in C programming language following a reverse communication spirit \citep{Metivier_2015_TOO}.
\end{itemize}
The objects in these types are coined \verb|acq|, \verb|sim|, \verb|fwi| and \verb|opt| for clarity.

Besides the data pointers listed above, the index and total number of the processors are global variables defined by \verb|iproc| and \verb|nproc|.  The header files are sitting in \texttt{SMIwiz/include}. The main part of the software is in \texttt{SMIwiz/src} and coded in C programming language to achieve high performance computing. We can launch \verb|make| to compile the code following the given \verb|Makefile|, which may be adapted depending on the user's environment and the path of the software packages. The compiled executable will be coined \verb|SMIwiz| in the folder \verb|SMIwiz/bin|. A number of unit tests, with the names of the folder \texttt{SMIwiz/run\_XXXX}, can be considered as the running templates to be adapted to fit the user's needs of modelling and imaging.

Similar to our 3D electromagnetic modelling and inversion software \verb|libEMMI| \citep{Yang_2023_libEMM,Yang_2023_3dcsem}, \verb|SMIwiz| automatically parses the argument list based on the parameter specified in the token. It checks out the value of each keyword after the equal sign \verb|=|. Multiple values for the same keyword are separated with comma. Different keywords will be separated using the space in the string. The code will automatically analyze  these keywords regardless the order of the input arguments.

All relevant parameters and the input files in binary or ASCII formats can be set up in a shell script, e.g., \verb|run.sh|. 
The first job \verb|run.sh| does is to write the input parameters in a text file \verb|inputpar.txt|. These parameters and input files will be easily adapted depending on the user's setup. 
Then, \verb|SMIwiz| is activated in \verb|run.sh| using MPI
\begin{verbatim}
mpirun -n 4 ../bin/SMIwiz $(cat inputpar.txt)
\end{verbatim}
where the number 4 indicates 4 shots of the seismic data are simultaneously simulated and/or inverted. To facilitate the job submission, the modelling and inversion will be launched via
\begin{verbatim}
bash run.sh
\end{verbatim}

\subsection{Parallelization}

\verb|SMIwiz| distributes jobs to different CPU cores in a bijective manner using MPI parallelization, so that the modelling for each shot are completely independent, therefore no collective communication among them. The numerical simulation of each shot has been parallelized using OpenMP thanks to local shared memory. Each processor will pick up their own source-receiver geometry from the same acquisition file. 

\verb|SMIwiz| has an argument \verb|shot_idx| to accept a list of shot indices so that specific shots will be modelled or utilized for imaging. For example, \verb|shot_idx=5,13,18,23| indicates shots 5,13,18,23 are modelled/imaged. The total number of MPI processors launched should be 4 to match the number of the shots in \verb|shot_idx|. If \verb|shot_idx| is not specified explicitly, the sources indexed from \verb|0001| to \verb|XXXX| will be consecutively handled by default, where \verb|XXXX| is the number of processors launched.

\begin{figure}
  \centering
  \includegraphics[width=0.9\linewidth]{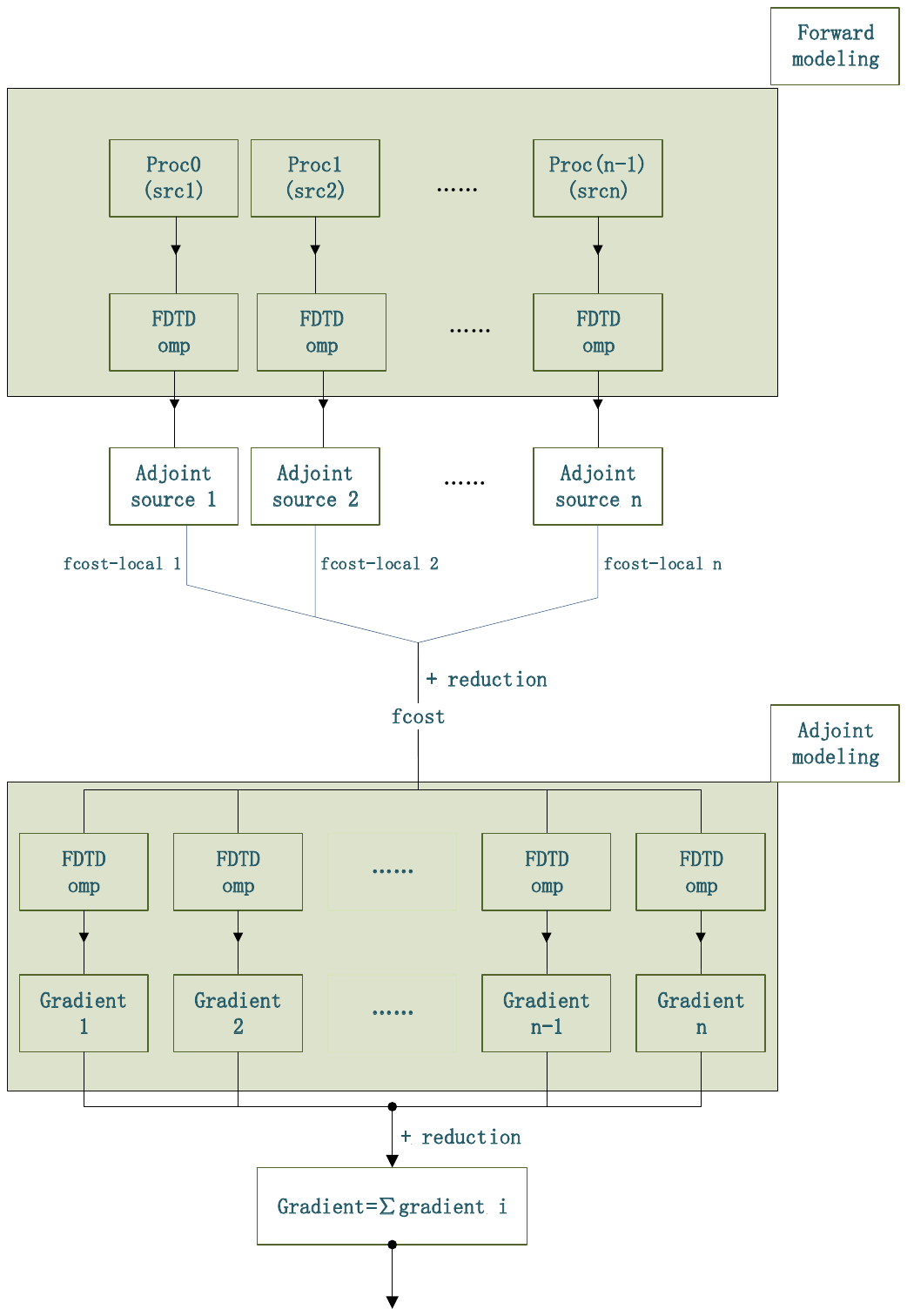}
  \caption{A workflow to build FWI gradient in \texttt{SMIwiz}}\label{fig:flowchart}
\end{figure}

In an imaging task such as RTM and FWI, we will perform a reduction to collect each piece of gradient/image information from different shots. By doing so, the amount of communication among processors is minimized to maximize the scalability of the imaging computations. The gradient building following this idea is shown in the flowchart in Figure~\ref{fig:flowchart}. 

\subsection{Running different jobs}

Depending on the value of \verb|mode| parameter, \verb|SMIwiz| runs different jobs. Setting \verb|mode=0| performs forward modelling based on the given density and velocity parameters. Assuming the observed data in binary are available, one can do FWI (\verb|mode=1|) or output the misfit gradients (\verb|mode=4|). When the seismic source is unknown, one may estimate the source signature before inversion via the method proposed in \cite{Pratt_1999_SWIb} (\verb|mode=5|).
Such a design eases the extension of the new functionalities with another different \verb|mode| value for more advanced processing.

\subsection{Acquisitions geometry}

To specify the survey layout, an acquisition files in ASCII format must be given to parameter \verb|acquifile=| to prescribe the source and receiver positions. With a text file \verb|acquifile=acqui.txt|, the source/receiver file is in the following form:
\begin{verbatim}
    z         x       y       azimuth    dip        src/rec(0/1)
   5.0        100.    0        0           0           0
   20.0       100.    0        0           0           1
   20.0       120.    0        0           0           1
   20.0       140.    0        0           0           1
   ...
   5.0        200.    0        0           0           0
   20.0       100.    0        0           0           1
   20.0       120.    0        0           0           1
   20.0       140.    0        0           0           1
   ...
\end{verbatim}
By default, the first columns gives the z coordinate, while the second and the third give the x and y coordinates. For a 2D simulation, the y coordinate will have no effect. At present, the space location \verb|(x,y,z)| in the unit of meters are active parameters, while the \verb|azimuth| and \verb|dip| parameters are not functional.
The last column is the flag of sources and receivers: 0 indicates a source and 1 indicates a receiver. After a source line, all receiver coordinates associated with this source will follow. Once a new source line comes, the following lines direct to the receivers recording waves from this new source. Since all acquisition geometry information will be stored in the same file, such an ASCII file can be large, typically for 3D configurations.

\subsection{Binary files for velocity, density and source time function}

\verb|SMIwiz| asks for three binary files for the detailed information on the medium properties  and the source time function, which can be specified through the following lines:
\begin{verbatim}
vpfile=vp
rhofile=rho
stffile=fricker
\end{verbatim}
where the velocity \verb|vp| and the density \verb|rho| of the subsurface must match the dimension of the simulation domain. The argument \verb|stffile| allows the user to work with their own source time function for modelling and inversion. In the above, \verb|fricker| is a Ricker wavelet. 

\subsection{Parameters in the simulation domain}

The size of the finite difference mesh will be specified in z, x and y directions with the dimension parameters \verb|n1|, \verb|n2| and \verb|n3|. The grid spacing are given by the arguments \verb|d1|, \verb|d2| and \verb|d3|. We add \verb|nb| PML layers on each side/face of the model. The simulation time can be determined using the number of time steps \verb|nt| and the temporal interval \verb|dt|. The simulation can be carried out with \verb|freesurf=1| or without free surface \verb|freesurf=0|. The central frequency of the wavelet must also be given. For example, a 3D modelling for 2000 time steps with stepsize $\Delta t=0.004$ s within a 3D mesh of size (\texttt{nx, ny, nz})=(81, 401, 401) and grid spacing $\Delta z=\Delta y=\Delta x=50$ m with free surface using a source with central frequency around 6 Hz in the shell script \verb|run.sh| can be configured via
\begin{verbatim}
freesurf=1 //free surface boundary condition
fm=6 //centre frequency of the wavelet
nt=2000 //number of time steps
dt=0.004 //temporal sampling
nb=20 //number of layers for absorbing boundaries
n1=81 //size of input FD model nz
n2=401 //size of input FD model nx
n3=401 //size of input FD model ny
d1=50 //grid spacing dz of the input FD model
d2=50 //grid spacing dx of the input FD model
d3=50 //grid spacing dy of the input FD model
\end{verbatim}
In case of a 2D simulation, the parameters \verb|n3| and \verb|d3| in the shell script can be removed.

\subsection{Inversion setup}

Our FWI uses bounded LBFGS algorithm. The number of nonlinear iterations are given by parameter \verb|niter|. At each iteration, it uses line search algorithm to determine the step length. The maximum number of possible line search can be specified by \verb|nls|. The memory length of LBFGS is given by \verb|npair|. By default, no preconditioning is applied (\verb|preco=0|). The depth preconditioner can be turned on by setting \verb|preco=1|, while the pseudo-Hessian preconditioner can be actived using \verb|preco=2|. By default, we activate bounds (\verb|bound=1|) with given minimum and maximum values for velocity (\verb|vpmin| and \verb|vpmax|) and density (\verb|rhomin| and \verb|rhomax|). The parametrizations of  $V_p-\rho$ and $V_p-I_p$ are referred to as family 1 and family 2, respectively. The index of parameters are given by \verb|idxpar|. The bounds for inversion parameters will be determined depending on the family of the parametrization considering the log transform of the physical parameters. The following lines set up an inversion of 50 iterations of preconditioned LBFGS algorithm to invert for $V_p$ and $I_p$: Every update are bounded by the minimum and maximum  wave speed 2000 m/s and 6500 m/s, as well as the minimum and maximum density 1000 kg/m$^3$ and 2800 kg/m$^3$.
\begin{verbatim}
niter=50 //number of iterations using LBFGS
nls=20 //number of line search per iteration
npair=5 //memory length in LBFGS
preco=1 //precondition

npar=2  //simultaneously invert 2 parameters - Vp and Ip
idxpar=1,2 //index of parameter to be inverted
bound=1 //bound the inversion parameters
vpmin=2000 //minimum value for wave speed
vpmax=6500 //maximum value for wave speed
rhomin=1000 //minimum value for density
rhomax=2800 //maximum value for density
\end{verbatim}

Note that \verb|npar=1| corresponds to monoparameter inversion while \verb|npar=2| corresponds to multiparmeter inversion. A binary file providing bathymetry information may be specified:
\begin{verbatim}
bathyfile=fbathy
\end{verbatim}
This file \verb|fbathy| should contain \texttt{n1*n2} values defining the depth of the seafloor. During the inversion, the gradient above the bathymetry will be set to 0 to avoid model update in the sea water. As the sources and receivers are normally placed above the seafloor, this mutes the strong acquisition footprint in the inverted model.

\subsection{Accuracy control}

The finite difference scheme of orders  $O(\Delta t^2, \Delta x^4)$ and  $O(\Delta t^2, \Delta x^8)$ can be specified by \verb|order=4| (default value) or \verb|order=8|. To ensure a consistent accuracy, the radius of Kaiser-windowed interpolation operator is set to be the same as half of the order  \verb|ri=order/2|, to ensure that the same number of points are used for interpolation and finite difference stencil. At the outset of numeric modelling, \verb|SMIwiz| always performs a sanity check on the CFL stability condition in equation \eqref{eq:cfl} and the dispersion requirement for sufficient points per wavelength. \verb|SMIwiz| will be automatically terminated if these conditions are not satisfied.

\subsection{Data weighting}

\begin{figure}
  \centering
  \includegraphics[width=\linewidth]{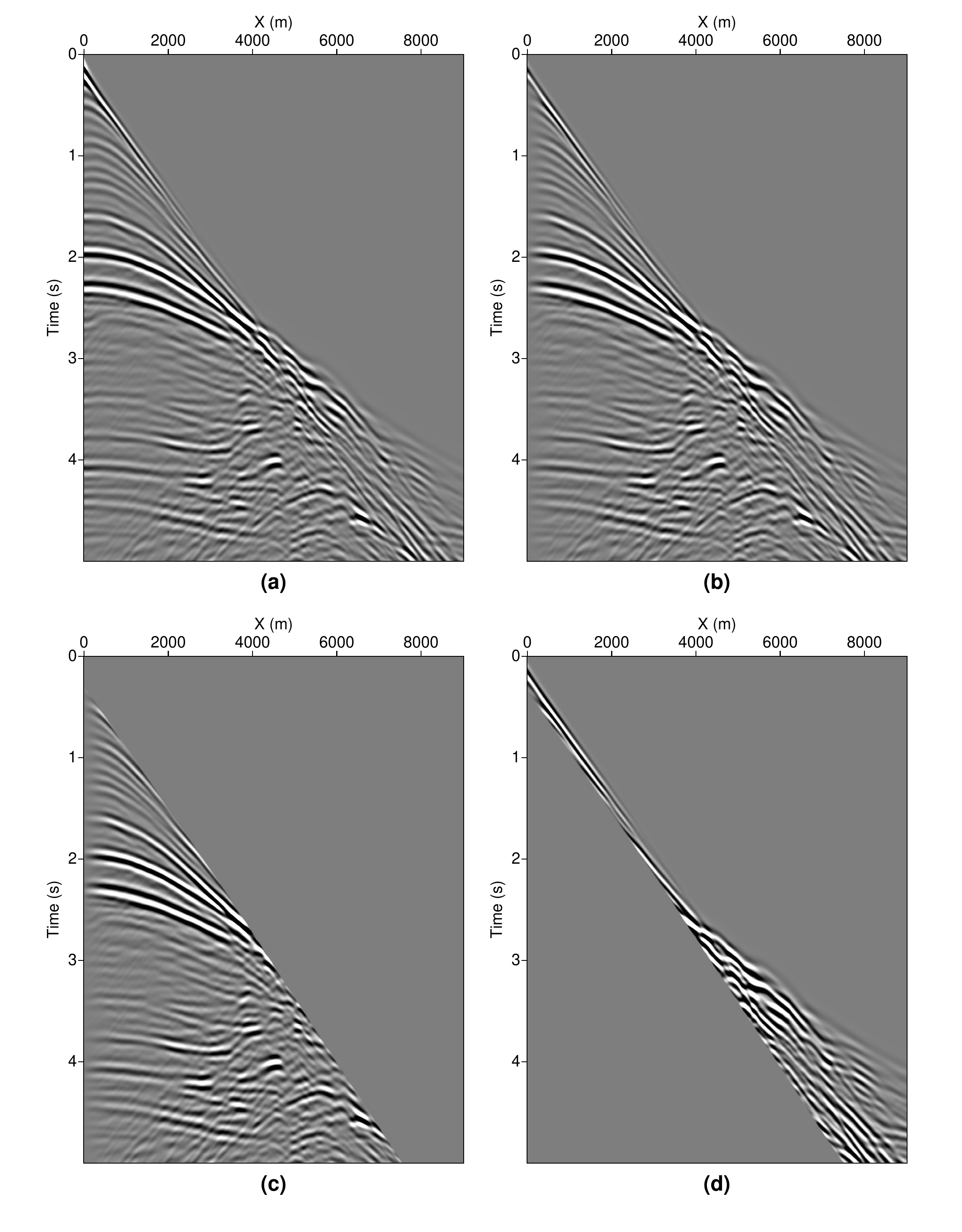}
  \caption{(a) A complete prestack seismic shot profile modelled from Marmousi model; (b) near offset data are tapered (\texttt{muteopt=0}, no muting); (c) front muting (\texttt{muteopt=1}), leaving only the reflections; (d) tail muting  (\texttt{muteopt=2}), leaving only the first arrivals and diving waves. }\label{fig:dataweight}
\end{figure}

We can apply different weights for data with varying offset. The parameter \texttt{dxwdat} allows us to configure intervals of offset sampling while different weights can be given by \texttt{xwdat}. As shown in Figure~\ref{fig:dataweight}a, a complete prestack shot profile has been modelled from Marmousi model.As illustrated in Figure~\ref{fig:dataweight}b, application of offset weighting leads to the dimming of near offset using 
\begin{verbatim}
dxwdat=200 //dx for data weighting
xwdat=0,0.3,0.7,1,1 //weights for dx increment
\end{verbatim}
RTM mainly uses reflections to retrieve high resolution structural image of subsurface. Such a purpose can be achieved by muting the front part of the data.
Since the main data driving the model update is the first arrivals, transmitted energy including direct waves and refractions are of particular importance in FWI. Figure~\ref{fig:dataweight}c and \ref{fig:dataweight}d examplifies the front and tail muting over the above shot profile. When muting is conducted, we use \verb|ntaper| points to have a  smoothing transition between muted and unmuted part of the data. The typical parameters for such data weighting look like
\begin{verbatim}
ntaper=20 //number of points for tapering
xmute1=0,3306,7534 //front mute line, x coordinate in m
tmute1=0.3,2.3,5.0 //front mute line, t coordinate in s
xmute2=0,3306,7534 //tail mute line, x coordinate in m
tmute2=0.3,2.3,5.0 //tail mute line, t coordinate in s
\end{verbatim}
The data weighting gives us great flexibility for preprocessing of the data in different imaging tasks.

\subsection{The output from \texttt{SMIwiz}}

The modelled data will be named as \texttt{dat\_XXXX} where \texttt{XXXX} is the index of the shots. The cross-correlation RTM image and normalized cross-correlation image will be \texttt{image\_xcorr} and \texttt{image\_normalized\_xcorr}. At each iteration, the updated model parameters and FWI gradient will be printed out, with names \texttt{param\_final} and \texttt{gradient\_final} (latest printing will overwrite existing file). In order to track the history of the inversion, all these intermediate gradients and updated parameters are also collected under the file names \texttt{param\_iter} and \texttt{gradient\_iter}. All these files are in binary format, implying that the user should be aware of the size of them to check out the results.

\section{Judicious implementation}\label{sec:implementation}

\begin{figure}
  \centering
  \includegraphics[width=\linewidth]{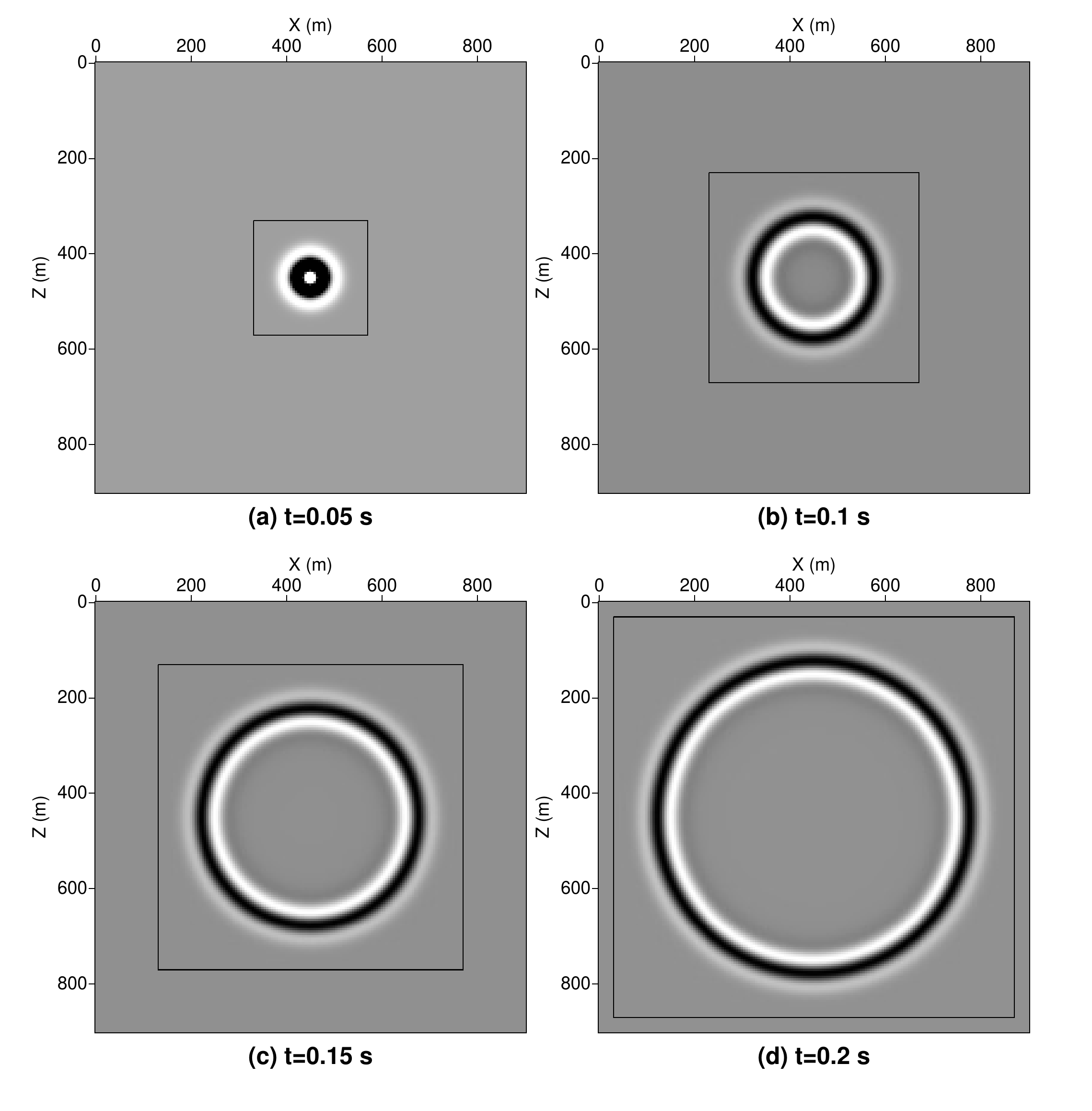}
  \caption{The size of computing box (marked by black box) expands with the evolution of the wavefield at each time step: (a) t=0.05 s; (b) t=0.1 s; (c) t=0.15 s; (d) t=0.2 s.}\label{fig:box}
\end{figure}

\subsection{Computing box}

\verb|SMIwiz| aims to do minimum amount of floating point operations to reach an intact wavefield simulation. To do that,  a cubic computing box which changes dynamically with the evolution of time has been designed. The minimum and maximum bounds for the grid index are determined based on the outermost wavefront, which is calculated based on the evolved number of time steps and maximum wave speed. In a forward modelling mode, the computing box is expanding. At the very beginning, the waves are confined in a small local cube, therefore requiring few gridpoints for finite difference; with the progress the modelling, the waves gradually spread into much large domain, therefore involving more gridpoints for modelling at each time step, as shown in Figure~\ref{fig:box}. In adjoint simulation, the computing box for the adjoint field is determined based on the outermost receivers, while the computing box for the forward field is shrinking backwards in time.

Let us remark that the efficiency of the application of computing box depends on where the seismic source is located. If the source is closer to one side of the model than other sides, the edge/face (in 2D/3D) of the computing box on this side will hit the boundary of the model first and remain unchanged. The worse case is that the source resides in the center of the model: in this case the edges/faces (2D/3D) of the computing box on each side extends. The efficiency gain due to computing box is more significant when the model is large. The amount of computation time might be only a fraction of the case without applying the box in 3D.

\subsection{Source wavefield reconstruction}

\begin{figure}
  \centering
  \includegraphics[width=\linewidth]{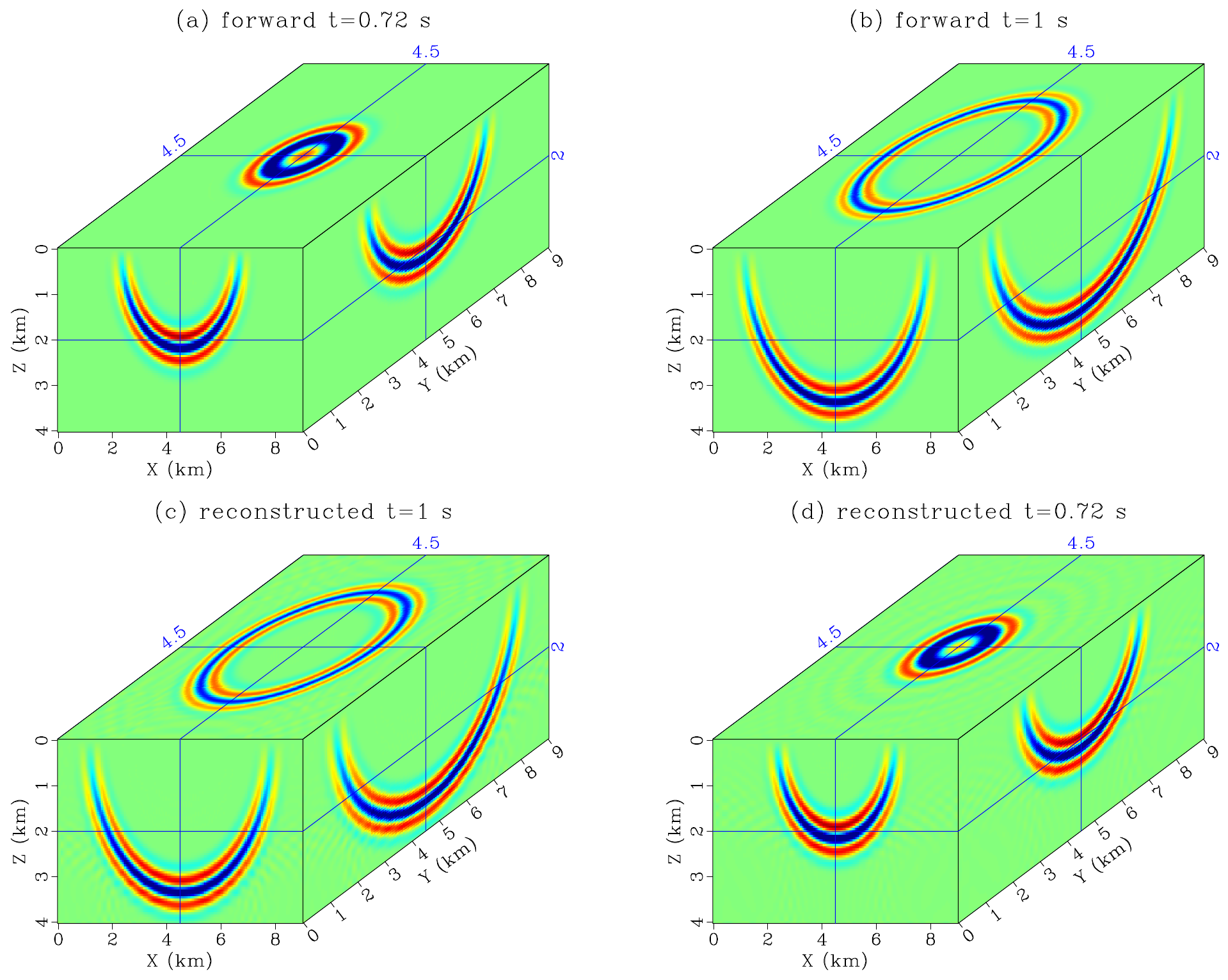}
  \caption{3D wavefield reconstruction in a constant medium with a source placed in the center of the model: The forward modelling lasts for 2 s with recorded wavefields at (a) t=0.72 s and (b) t=1 s, then backward reconstructed from 2 s to 0 s, with recorded wavefields at (c) t=1 s and (d) t=0.72 s. }\label{fig:waverec}
\end{figure}

A common significant issue in nonlinear FWI and RTM is the construction of an inversion gradient or an image, which requires simultaneously accessing forward and adjoint fields at every time step, cf. a detailed analysis of this problem in \cite{Yang_2016_CAR}. Since the adjoint field has an opposite time order compared to forward field while storing all snapshots of the forward field at every time leads to huge amount of memory consumption and heavy IO traffic, the source wavefield is reconstructed by interpolating the decimated boundaries in backward time order. To drastically reduce the memory consumption, the boundaries of forward field is only stored at every \texttt{dr} time step, where the ratio \texttt{dr} depends on the velocity contrast with minimum value
\begin{equation}
  \texttt{dr}=\frac{5 V_{p\max} }{V_{p\min}}.
\end{equation}
Note that such an estimate is made based on 4-th order FD scheme in space, while the same estimation is approximately valid for 8th order FD as well, see \cite[equation 4]{Yang_2016_WRB}.
Assume $V_{p\max}\ge 2V_{p\min}$, the default decimation rate is set to be \texttt{dr=10} in 3D.  Since there is sufficient amount of memory available in 2D case, the default value is \texttt{dr=1} allowing all boundaries at every timestep stored in memory so that the incident wavefield can be perfectly reconstructed backwards in time.
The method was initially validated in 2D case \citep{Yang_2016_WRB}. Despite the idea is no more new,  combining decimation and interpolation for significant memory reduction in wavefield reconstruction is critical to work with 3D imaging applications.
Here we present the first numerical demonstration of this method in 3D:  As shown in Figure~\ref{fig:waverec}, a precise reconstruction of the source field in backward time order is feasible, despite the fact that only 1 sample out of 10 time steps of the boundaries is stored).

\subsection{Batchwise job scheduling}

Up to now, a number of features possessed by \verb|SMIwiz| assumes sufficient number of computer memory and processors, so that seismic shots are processed in bijective fashion. For 3D large scale imaging applications, this can be done as long as a modern cluster is at hand. It becomes challenging to do given only a small workstation. This in fact can still be done in \verb|SMIwiz| thanks to a batchwise job scheduling strategy. Each time, we submit multiple modelling/inversion jobs as a batch, corresponding to a small subset of the target sources. For RTM, we can sequentially loop over the observed seismic data, as all images associated with different sources should be summed together.

However, catching a subset of the shots close to each other in FWI may introduce a strong bias when computing the misfit gradient, which will finally drive the inversion towards an unexpected direction. To get rid of this bias, we introduce the so-called stochastic gradient descent (SGD) method. There is no need to do any modification in the code of \verb|SMIwiz|. Instead, we design a random shuffling algorithm in the shell script to submit inversion jobs in batches. Each batch is responsible for one iteration of steepest descent update of the model, using a specific number of shots determined randomly. Starting from the given initial model, the updated model at each batch will be used as the initial guess for next batch.

Assume there are only 50 CPU cores available in a small workstation while the observed data has 400 shots. Direct submitting 400 shots to 50 processors is infeasible for \verb|SMIwiz| to do FWI/RTM  in this case. The following lines implement the batchwise SGD algorithm by first writing the indices (generated via random shuffling algorithm) of the shots into a file, then update the model parameters batch by batch according to the shot indices prescribed in the file. Note that the number of shots do not need to be a multiple of the batchsize. The following code will schedule the last batch according to its actual size.

\begin{verbatim}
batchsize=50 #each group has less than or equal to 50 shots
ntotal=400 #total number of shots used in the inversion
j=0
k=0
for i in `shuf -i 1-$ntotal -n $ntotal`; do
    if [ `expr $j % $batchsize` -eq 0 ]; then 
        k=$[ $k + 1 ]
	echo 'batch' $k
    	echo $i > batch$k #write into a new file
    else
    	echo $i >> batch$k #append into an existing file
    fi
    j=$[$j + 1]
    echo $i
done
nbatch=$k

for k in `seq 1 $nbatch`; do
    echo "#input parameters ...
" >inputpar$k.txt
    shots=`echo $(echo $(cat batch$k)) | tr ' ' ','`
    echo "shots="$shots >> inputpar$k.txt
    if [ $k -eq 1 ]; then
	echo "vpfile=vp_init">>inputpar$k.txt
    else #use updated vp as initial model
	echo "vpfile=param_final">>inputpar$k.txt 
    fi

    j=0 #count the actual number of shots in this batch
    for i in $(cat batch$k); do
	j=$[$j + 1]
    done
    echo batch$k, 'nshot='$j, 'shot_idx='$shots
    mpirun -n $j ../bin/SMIwiz $(cat inputpar$k.txt)
done
\end{verbatim}
In the above script, the keyword \verb|shot_idx| is the key parameter to accept the indices of the shots from random selection.

\section{Numerical examples}

\subsection{Nonlinear multiparameter inversion}

\begin{figure}
  \centering
  \includegraphics[width=0.95\linewidth]{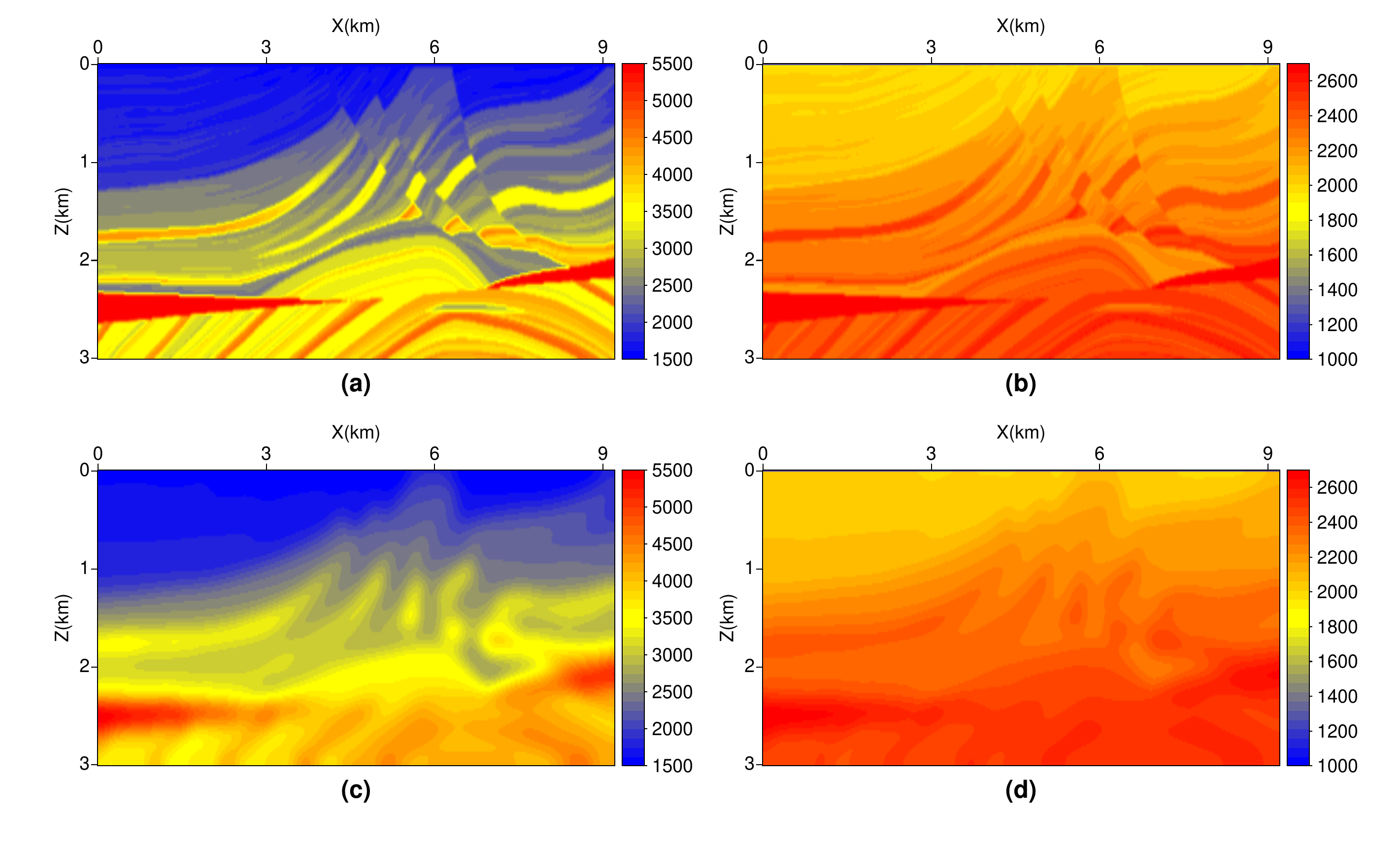}
  \caption{(a) True Marmousi velocity model; (b) Density model derived from true velocity model using Gardner's law; (c) Smoothed velocity as the starting model; (d) Density derived from smoothed velocity model using Gardner's law.}\label{fig:marmousi}
\end{figure}

We begin our demonstration of \verb|SMIwiz| by nonlinear multiparameter inversion of wave speed and density in classic Marmousi model (3 km $\times$ 9.2 km). The model is meshed into $(\texttt{nz}, \texttt{nx})$= (151, 461) samples with grid spacing $\Delta x=\Delta z=20$ m. The source for this experiment is a Ricker wavelet with peak frequency 5 Hz. Starting velocity model is obtained by moderate smoothing of the true model; both the true and initial density models are derived from the velocity using Gardner's law, as shown in Figure~\ref{fig:marmousi}. In total, 24 shots at 5 m depth  are used in the inversion. We consider a fixed spead surface acquisition geometry: 451 receivers with equal spacing are deployed evenly at 20 m depth to record seismic data excited by each shot.

\begin{figure}
  \centering
  \includegraphics[width=0.95\linewidth]{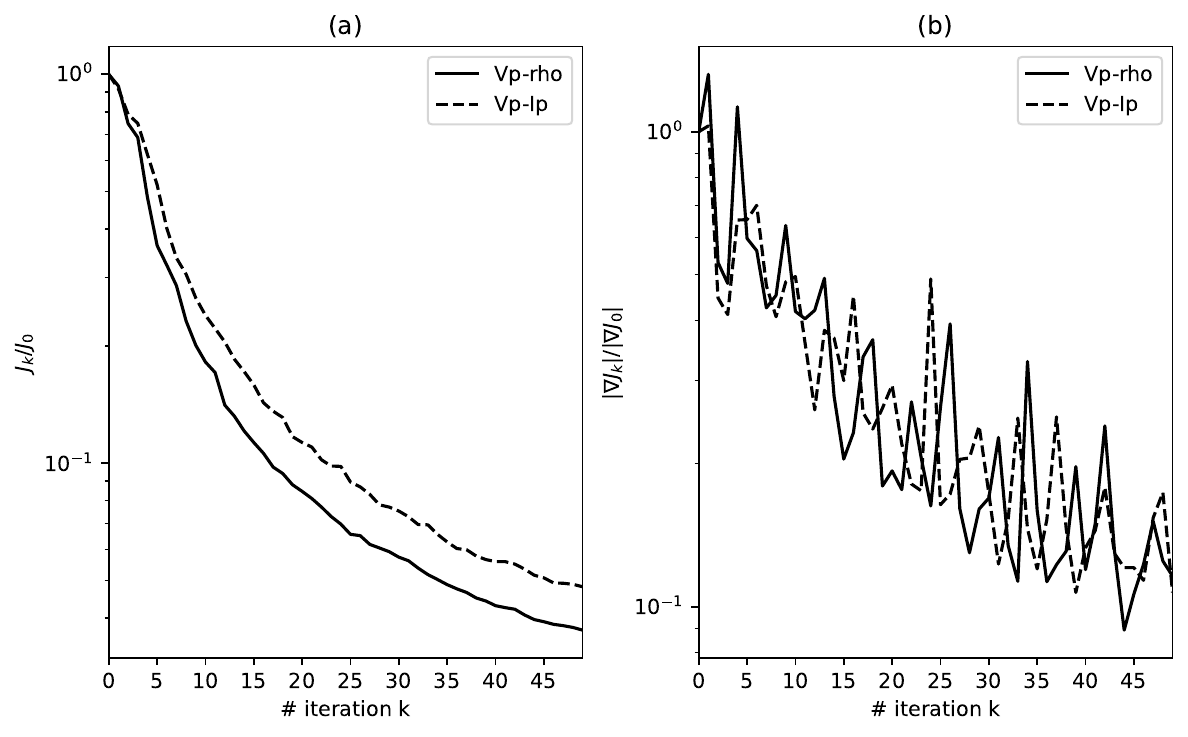}
  \caption{Convergence history of (a) normalized misfit and (b) gradient for two-parameter inversion under family 1 nad family 2.}\label{fig:marmconv}
\end{figure}

We perform multiparameter inversion for the Marmousi model in both family 1 ($V_p-\rho$ parametrization) and family 2 ($V_p-I_p$ parametrization) using 50 iterations of preconditioned LBFGS algorithm.
According to the normalized misfit and the norm of the gradient Figure~\ref{fig:marmconv}, the inversion under in family 1 and family 2 converges very well.
Within limited number of iterations,  a slightly faster convergence rate and lower final misfit value in $V_p-\rho$ parametrization have been reached in Figure~\ref{fig:marmconv}a. 

\begin{figure}
  \centering
  \includegraphics[width=0.95\linewidth]{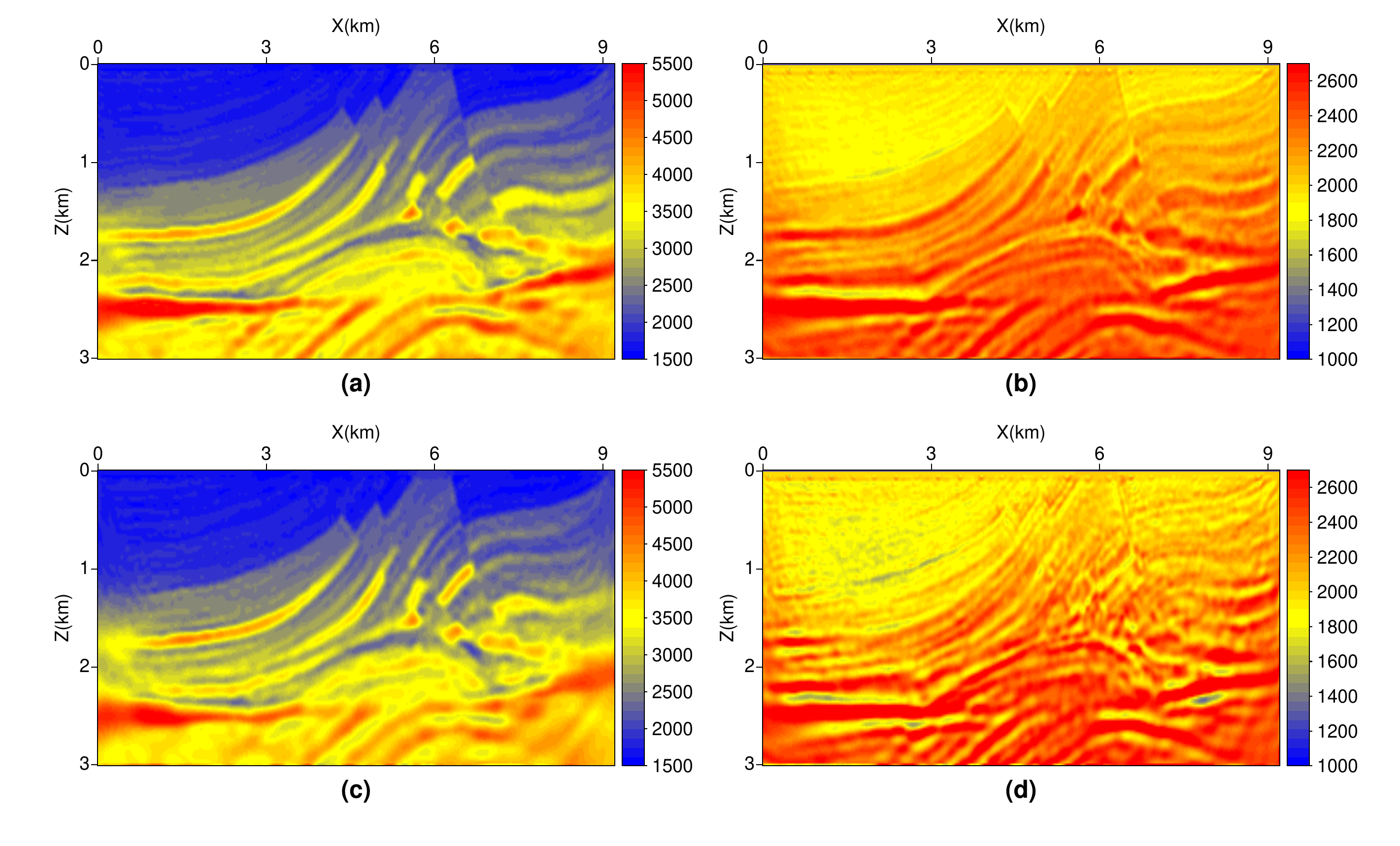}
  \caption{Inverted  velocity (a, c) and density (b, d) under parametrizations $V_p-\rho$ (a, b) and $V_p-I_p$ (c, d). The density under $V_p-I_p$ is not a direct inversion parameter, and is derived by $\rho=I_p/V_p$.}\label{fig:marminv}
\end{figure}

As can be seen from Figure~\ref{fig:marminv}, the inverted velocity and density under $V_p-\rho$ parametrization are closer to the true model, compared with the ones obtained under $V_p-I_p$ parametrization. The inverted density panel under $V_p-I_p$ parametrization presents more high wavenumber oscillations than the one under $V_p-\rho$.  This reveals the difference and importance of the choice of an appropriate model parametrization when performing multiparamter inversion.
The above result should be understandable based on a radiation pattern analysis, see \cite[Figure 1]{Yang_2018_TRN}.

\subsection{Reverse time migration}

\begin{figure}
  \centering
  \includegraphics[width=\linewidth]{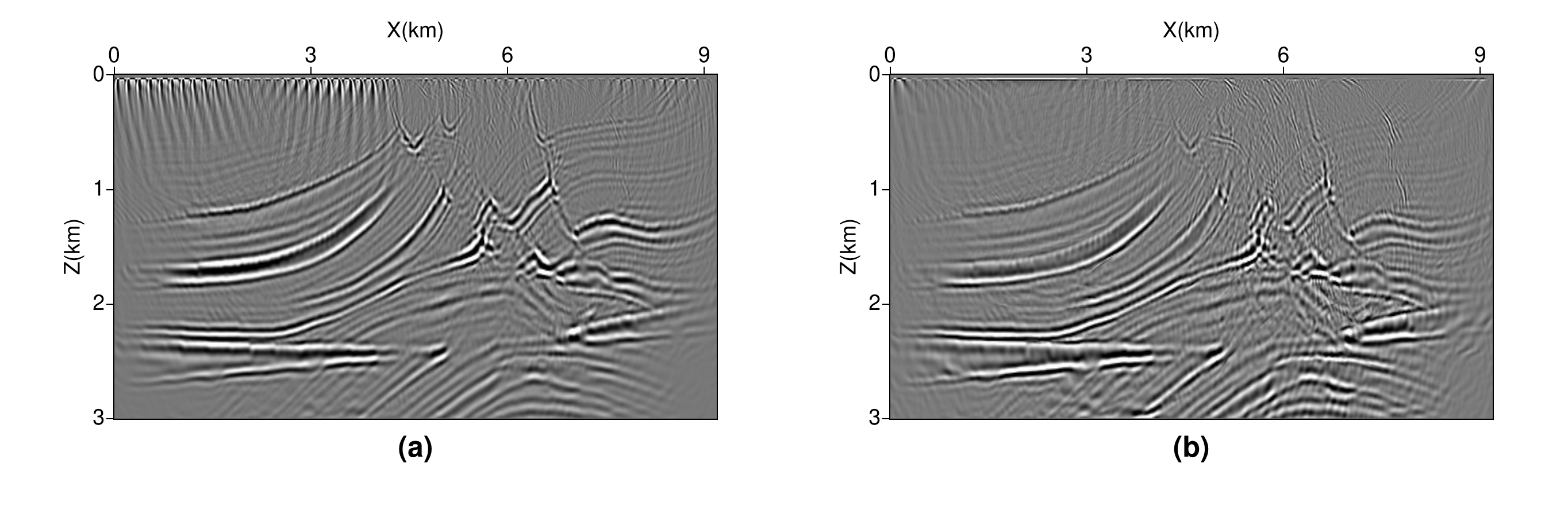}
  \caption{RTM image using (a) cross-correlation imaging condition and (b) normalized cross-correlation imaging condition}\label{fig:rtm}
\end{figure}

Once a reasonably accurate velocity model is obtained, we can then perform RTM imaging. Here we start from a resampled Marmousi model for RTM test with the grid spacing
$\Delta x=\Delta z=12$ m and the size $(\texttt{nz}, \texttt{nx})$=(251, 767). The reason for refined sampling is that RTM generally runs at much higher frequency than FWI (demanding more points per wavelength according to dispersion requirement), to accurately retrieve the interface locations as a structural image. Figure~\ref{fig:rtm}a and \ref{fig:rtm}b display the RTM images computed based on cross-correlation imaging condition and normalized cross-correlation imaging condition. The migration artifacts in the shallow part of Figure~\ref{fig:rtm}a has been greatly suppressed in Figure~\ref{fig:rtm}b by introducing source field normalization; meanwhile, the interface at depth constructed by seismic interference become much more energetic and sharper.

\subsection{3D FWI of Overthrust model}

\begin{figure}
  \centering
  \includegraphics[width=\linewidth]{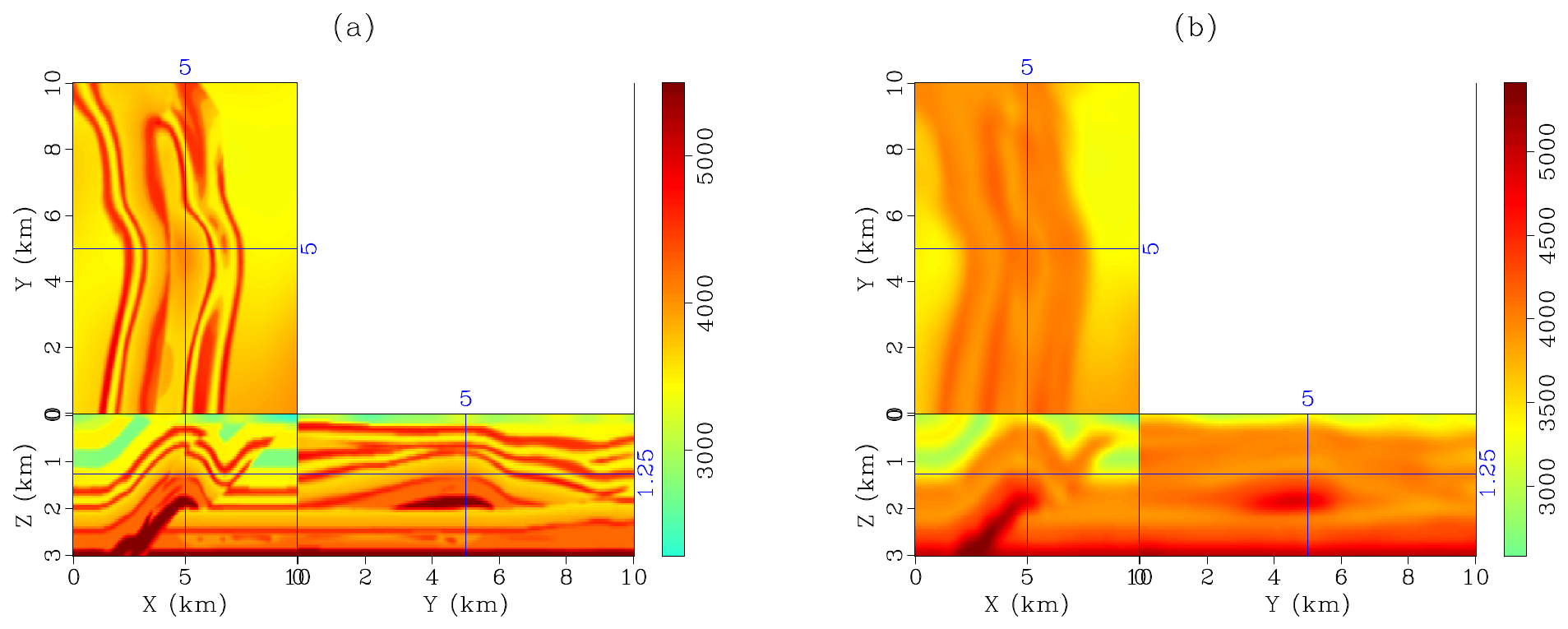}
  \caption{(a) Part of 3D Overthrust velocity model; (b) Smoothed version of (a).}\label{fig:overthrust}
\end{figure}

The third example is a 3D FWI test on Overthrust model with interesting geological features such as anticline and the dipping faults. As shown in Figure~\ref{fig:overthrust}a, the  selected part of this 3D model spans over $10\times 10$ km$^2$  surface area, extending up to 3 km in depth. 
The model is uniformly sampled along x-, y- and z- directions: $\Delta x=\Delta y=\Delta z=50$ m.
The initial model in Figure~\ref{fig:overthrust}b is a strongly smoothed version of the true model. In this example, the density model is again created using Gardner's law, and kept constant. We focus here on monoparameter inversion to retrieve the velocity since it is the first order parameter to explain the phase and amplitude responses of the seismic data. The central frequency of this inversion is around 6 Hz. The modelling lasts for 1500 steps with temporal sampling $\Delta t=0.004$ s.

\begin{figure}
  \centering
  \includegraphics[width=0.7\linewidth]{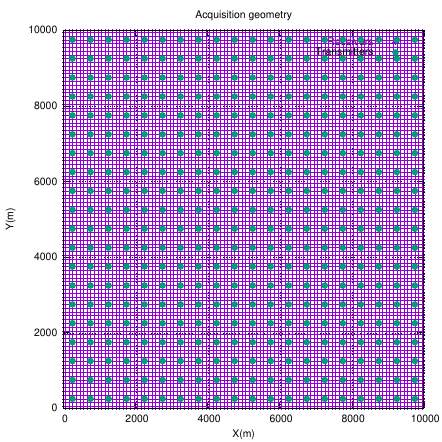}
  \caption{The survey layout sheet of sources (red crossings) and the receivers (green dots)}\label{fig:survey}
\end{figure}

As illustrated in Figure~\ref{fig:survey}, we deployed 400 shots evenly distributed in horizontal plane at the depth of 5 m. For each source, we use 10000 receivers  at the depth of 10 m to record the seismic data. The source and receiver separations are 500 m and 100 m respectively. Such a configuration of the source-receiver geometry is to mimic the real acquisition performed in the field: there are tens or hundreds of
thousands of explosive sources excited while thousands of receivers are deployed to record the wavefields. The reciprocity is often used to switch the role of source and receivers to reduce the computational cost for 3D FWI of the field data, so the actual shots in FWI are common receiver gathers.

\begin{figure}
  \centering
  \includegraphics[width=0.7\linewidth]{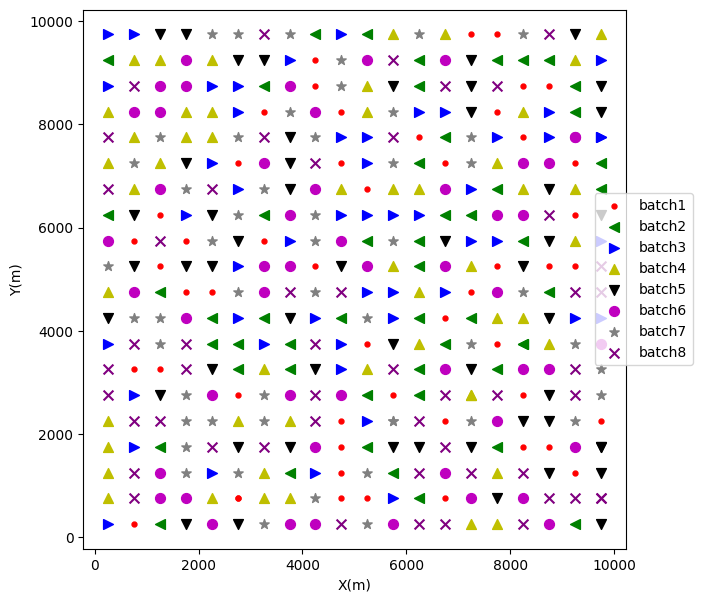}
  \caption{Randomly selected shots for 8 batches of stochastic gradient descent}\label{fig:randomshots}
\end{figure}

\begin{figure}
  \centering
  \includegraphics[width=0.84\linewidth]{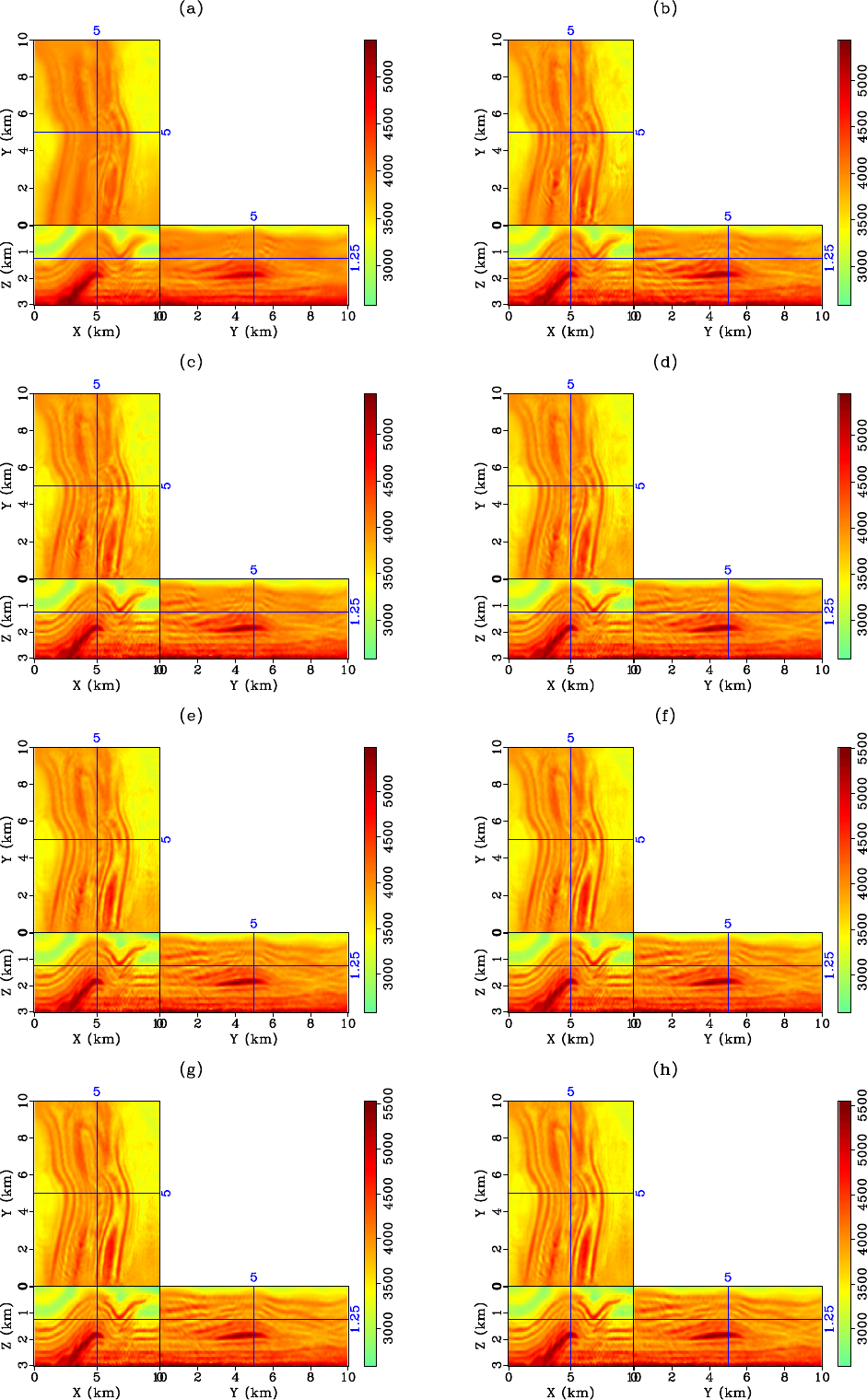}
  \caption{The update velocity model at (a) batch 1; (b) batch 2; (c) batch 3; (d) batch 4; (e) batch 5; (f) batch 6; (g) batch 7; (h) batch 8.}\label{fig:vpinv}
\end{figure}

  Taking into acount 20 PML layers on each side, this model has a grid size of $(\texttt{nz}, \texttt{nx}, \texttt{ny})$=(101, 241, 241).
The size of this model and the data set indeed poses a significant challenge for FWI algorithms. To handle such a large scale problem on our small workstation, we employed SGD algorithm to divide all 400 shots into 8 batches, resulting in 50 shots per batch. Figure~\ref{fig:randomshots} shows the pattern of the random selection algorithm at each batch using different markers: each time the randomly selected shots do not overlap with other batches. This implies all the data will be touched only once during the inversion. In each batch, we limit the number of iterations to only 3. Figure~\ref{fig:vpinv} displays all the updated velocity model after each batch of FWI: the evolution of the model clearly demonstrates the improvement of the resolution. Of course, the resolution can be even higher by repeating more batchwise FWI inversion, which is not pursued since the main feature of the model has been well retrieved.

\section{Memory consumption and computational efficiency}

Let us emphasize that all the computations were done using only a small workstation: it runns with Intel(R) Xeon(R) Gold 6258R CPU @2.7GHz, possessing 28 sockets with 2 cores per socket.  When simultaneously launching 50 processes, each shot only consumes 1.5 GB memory. Therefore, it is definitely a remarkable achievement to accomplish such a large size 3D FWI problem using so little computing resources.

\begin{figure}
  \centering
  \includegraphics[width=0.9\linewidth]{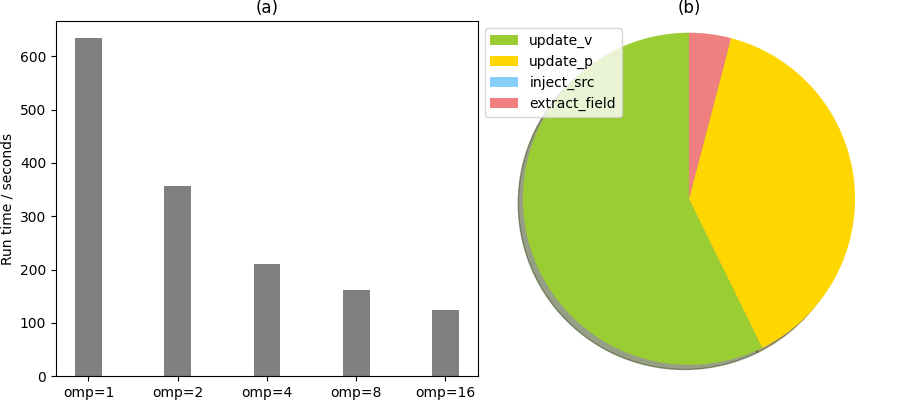}
  \caption{(a) The runtime of our modelling decreases when increasing the number of OpenMP threads; (b) the ratio of the runtime for updating particle velocities, pressure, injecting source and extracting wavefield in one modelling}\label{fig:runtime}
\end{figure}

Note that building each gradient/imaging condition requires 3 times of modelling, a forward modelling, a backward reconstruction of the forward field and an adjoint modelling. Since we map the modelling shot by shot in a bijective manner, the computing cost will  scale with total number of shots linearly. Consequently, we can make a direct accessment on the efficiency of \verb|SMIwiz| by checking the running time for one modelling job. We now examine the performance of our implementation for one shot simulation in 3D Overthrust model. Figure~\ref{fig:runtime}a shows  that double the number of OpenMP threads halves the runtime at the onset, but the efficiency gained by increasing the number of OpenMP threads quickly goes down when the number of OpenMP threads is larger than 4. Figure~\ref{fig:runtime}b illustrates that updating particle velocities and pressure takes most of the computing time, while the time cost for injecting source is almost negligible. Because the number of receivers (10000) is pretty large, the extraction of the wavefield at receiver positions also takes a fraction of total runtime.

\section{Discussion}

The acoustic wave equation can also be presented in the 2nd order order form by eliminating the particle velocities in equation \eqref{eq:acoustic}. We prefer the 1st order system because (1) PML boundaries are much easier to implement in 1st order system than 2nd order one, which can be much more efficient to absorbing the artificial reflections than sponge \citep{Cerjan_1985_NBC} and Clayton-Enquist \citep{Clayton_1977_ABC} absorbing boundary conditions in truncated mesh; (2) the impact of density can be explicitly expressed in front of fields, which simplifies the derivation of inversion gradient with respect to density; (3) an accurate injection of different source types can be conveniently injected close to the free surface to extend the functionality of \verb|SMIwiz| in the future.

At present, only basic C codes  are included in form the mainbody of SMIwiz software. The efficiency of the modelling engine in \verb|SMIwiz| will be further improved using GPU-based hardware acceleration. Since programming with CUDA involves expert knowledge, GPU codes for modelling are intentionally excluded in this package.

Domain decomposition has not been considered in our implementation for some reasons. The memory over one computer node in modern cluster is normally greater than 32 or 64 GB. Processing 3D imaging using our wavefield reconstruction requires only 1.5 GB of the computer memory for 1500 time steps. Even increasing the number of time steps higher than one order of magnitude can be completed using one computer node. It is therefore not necessary to do domain decomposition for acoustic applications in low frequency regime. However, domain decomposition for huge 3D simulation in elastic case is definitely required. Typically, the mesh size becomes extremely large if the domain is partitioned using very small grid spacing. This is the case if RTM and FWI must be performed at higher frequency band.   We refer to    \cite{minkoff2002spatial} on domain decomposition for more details.

When there are sufficient number of cores and memory available, one can directly submit all shots/data for standard FWI and RTM imaging without the need of batchwise job submission. An example script \verb|run_slurm.sh| for submitting parallel jobs on cluster using \verb|slurm| have been included in \verb|SMIwiz/run_fwi2d|. Processing in batches is an effective workaround only when we are limitted by computational resources. The batchwise job scheduling used for 3D FWI applies also to 3D RTM equally. Since the computational procedure of RTM is similar to building FWI gradient, there is no need to repeat the same strategy in our numerical demonstration. Both of them are based source field reconstruction strategy.

Instead of reconstructing the source field backwards in time, one may construct the FWI gradient by a mixed domain approach: The modelling of the forward and adjoint wavefields is in the time domain, but their frequency domain counterparts are integrated on the fly using discrete Fourier transform \citep{Nihei_2007_FRM,Sirgue_2007_SM3D}. The FWI gradient can then be computed by multiplication between frequency domain forward and adjoint wavefields. Such a method avoids one modelling in building the FWI gradient, but requires additional computer memory to store the wavefields for every discrete frequency. Unfortunately, this method becomes extraordinarily memory demanding for RTM applications, in which a broad range of frequencies are needed to form a high resolution migration image.

A notoriously difficult problem is the FWI cycle skipping issue. The optimization is often trapped into local minima when FWI is performed from a very poor starting model, using the data without long offset, low-frequency information. A number of solutions to mitigate such difficulties have been proposed, using multiscale strategies from low to high frequencies \citep{Bunks_1995_MSW}, layer stripping from long to near offset \citep{bian2011layer}, the modifications of the misfit functions based on cross-correlation \citep{Luo_1991_WETc}, deconvolution \citep{Warner_2016_AWI} or optimal transport distance \citep{Metivier_2016_OTI}. These ideas can be explored easily within the framework of \verb|SMIwiz|, but out of the scope of this work. We aim to provide a public platform rather than completing all the techniques in the literature.

\section{Conclusions}

\verb|SMIwiz| serves as a consistent realization of seismic modelling and imaging. It has a mechanism to do 2D/3D imaging flexibly, with some data weighting and preprocessing capabilities.
\begin{itemize}
\item The idea of computing box is simple yet powerful to restrict the modelling in a local cubic grid to avoid redundant computations, which significantly speeds up the forward and adjoint modelling.

\item To efficiently compute RTM image and the inversion gradient, we have implemented wavefield reconstruction using decimated boundaries, so that the source field can be accurately retrieved backwards in time using interpolated boundaries. It is indeed the key factor for \verb|SMIwiz| to resolve practical 3D large scale imaging problems.

\item The batchwise job scheduling gives flexibility to the user to take advantages of the limited computer memory and processors. This can be further explored by the user as long as there are sufficient computer resources accessible in a modern cluster.
\end{itemize}

We have demonstrated the versatile modelling and imaging functionalities of \verb|SMIwiz| selectively. There are other interesting modules still under development, for example, linearized waveform inversion via Born approximation. The module of modelling has been made self-contained so that it can be extended into elastic regime incorporating anisotropy and attenuation, while the backbone of nonlinear optimization has all necessary components to extend from quasi-Newton approach into a Newton-CG minimization.

\section*{Acknowledgements}

The author receives financial support from National Natural Science Foundation of China (42274156). Many user-friendly features of this software benefits from the extensive development experience with \verb|Madagascar| open software package. The design of \verb|SMIwiz| takes the benefit of \verb|ToyxDAC_Time|, which is a Fortran code for 2D/3D seismic modelling and inversion that the author works on during his postdoctral research. The author thanks the kind supervision and motivating discussions with Romain Brossier, Ludovic M\'etivier and Jean Virieux during his stay in SEISCOPE consortium in France.

\section*{Data Availability}
The data used for the research has been included in the software package - \verb|SMIWiz|.

\section*{Declaration of Competing Interest}

I declare that I have no financial and personal relationships with other people or organizations that can inappropriately influence our work.

\bibliographystyle{apalike}

\begin{thebibliography}{}

\bibitem[Abdelkhalek et~al., 2009]{abdelkhalek2009fast}
Abdelkhalek, R., Calandra, H., Coulaud, O., Roman, J., and Latu, G. (2009).
\newblock {Fast seismic modeling and reverse time migration on a GPU cluster}.
\newblock In {\em 2009 International Conference on High Performance Computing
  \& Simulation}, pages 36--43. IEEE.

\bibitem[Baysal et~al., 1983]{Baysal_1983_RTM}
Baysal, E., Kosloff, D., and Sherwood, J. (1983).
\newblock Reverse time migration.
\newblock {\em Geophysics}, 48:1514--1524.

\bibitem[Bian and Yu, 2011]{bian2011layer}
Bian, A. and Yu, W. (2011).
\newblock Layer-stripping full waveform inversion with damped seismic
  reflection data.
\newblock {\em Journal of Earth Science}, 22(2):241--249.

\bibitem[Bohlen, 2002]{Bohlen_2002_PVF}
Bohlen, T. (2002).
\newblock Parallel 3-{D} viscoelastic finite-difference seismic modeling.
\newblock {\em Computers \& Geosciences}, 28:887--899.

\bibitem[Bohlen et~al., 2015]{bohlen2015sofi3d}
Bohlen, T., Nil, D.~D., and Jetschny, D. K.~S. (2015).
\newblock {SOFI3D - seismic modeling with ﬁnite diﬀerences 3D - acoustic
  and viscoelastic version - user guide}.

\bibitem[Brossier, 2011]{Brossier_2011_TDF}
Brossier, R. (2011).
\newblock Two-dimensional frequency-domain visco-elastic full waveform
  inversion: Parallel algorithms, optimization and performance.
\newblock {\em Computers \& Geosciences}, 37(4):444 -- 455.

\bibitem[Bunks et~al., 1995]{Bunks_1995_MSW}
Bunks, C., Salek, F.~M., Zaleski, S., and Chavent, G. (1995).
\newblock Multiscale seismic waveform inversion.
\newblock {\em Geophysics}, 60(5):1457--1473.

\bibitem[Carcione, 2010]{Carcione_2010_GFP}
Carcione, J.~M. (2010).
\newblock A generalization of the fourier pseudospectral method.
\newblock {\em Geophysics}, 75(6):A53--A56.

\bibitem[Cerjan et~al., 1985]{Cerjan_1985_NBC}
Cerjan, C., Kosloff, D., Kosloff, R., and Reshef, M. (1985).
\newblock A nonreflecting boundary condition for discrete acoustic and elastic
  wave equations.
\newblock {\em Geophysics}, 50(4):2117--2131.

\bibitem[Clayton and Engquist, 1977]{Clayton_1977_ABC}
Clayton, R. and Engquist, B. (1977).
\newblock Absorbing boundary conditions for acoustic and elastic wave
  equations.
\newblock {\em Bulletin of the Seismological Society of America},
  67:1529--1540.

\bibitem[Dumbser and K{\"a}ser, 2006]{dumbser2006arbitrary}
Dumbser, M. and K{\"a}ser, M. (2006).
\newblock {An arbitrary high-order discontinuous Galerkin method for elastic
  waves on unstructured meshes—II. The three-dimensional isotropic case}.
\newblock {\em Geophysical Journal International}, 167(1):319--336.

\bibitem[Etgen et~al., 2009]{Etgen_2009_ODI}
Etgen, J., Gray, S.~H., and Zhang, Y. (2009).
\newblock An overview of depth imaging in exploration geophysics.
\newblock {\em Geophysics}, 74(6):WCA5--WCA17.

\bibitem[Fabien-Ouellet et~al., 2017]{fabien2017time}
Fabien-Ouellet, G., Gloaguen, E., and Giroux, B. (2017).
\newblock Time-domain seismic modeling in viscoelastic media for full waveform
  inversion on heterogeneous computing platforms with opencl.
\newblock {\em Computers \& Geosciences}, 100:142--155.

\bibitem[Fornberg, 1998]{Fornberg_1998_NFD}
Fornberg, B. (1998).
\newblock Classroom note: Calculation of weights in finite difference formulas.
\newblock {\em SIAM review}, 40(3):685--691.

\bibitem[Graves, 1996]{Graves_1996_SSW}
Graves, R. (1996).
\newblock Simulating seismic wave propagation in 3{D} elastic media using
  staggered-grid finite differences.
\newblock {\em Bulletin of the Seismological Society of America},
  86:1091--1106.

\bibitem[Hicks, 2002]{Hicks_2002_ASR}
Hicks, G.~J. (2002).
\newblock Arbitrary source and receiver positioning in finite-difference
  schemes using {K}aiser windowed sinc functions.
\newblock {\em Geophysics}, 67:156--166.

\bibitem[Hu et~al., 1999]{hu1999analysis}
Hu, F.~Q., Hussaini, M.~Y., and Rasetarinera, P. (1999).
\newblock An analysis of the discontinuous galerkin method for wave propagation
  problems.
\newblock {\em Journal of Computational Physics}, 151(2):921--946.

\bibitem[K{\"o}hn et~al., 2012]{Kohn_2012_IMP}
K{\"o}hn, D., De~Nil, D., Kurzmann, A., Przebindowska, A., and Bohlen, T.
  (2012).
\newblock On the influence of model parametrization in elastic full waveform
  tomography.
\newblock {\em Geophysical Journal International}, 191:325--345.

\bibitem[Komatitsch and Martin, 2007]{Komatitsch_2007_GEO}
Komatitsch, D. and Martin, R. (2007).
\newblock {An unsplit convolutional perfectly matched layer improved at grazing
  incidence for the seismic wave equation}.
\newblock {\em Geophysics}, 72(5):SM155--SM167.

\bibitem[Komatitsch and Vilotte, 1998]{Komatitsch_1998_SEM}
Komatitsch, D. and Vilotte, J.~P. (1998).
\newblock The spectral element method: an efficient tool to simulate the
  seismic response of 2{D} and 3{D} geological structures.
\newblock {\em Bulletin of the Seismological Society of America}, 88:368--392.

\bibitem[Lailly, 1983]{Lailly_1983_SIP}
Lailly, P. (1983).
\newblock The seismic inverse problem as a sequence of before stack migrations.
\newblock In Bednar, R. and Weglein, editors, {\em Conference on {I}nverse
  {S}cattering, Theory and application, Society for Industrial and Applied
  Mathematics, Philadelphia}, pages 206--220.

\bibitem[Levander, 1988]{Levander_1988_FOF}
Levander, A.~R. (1988).
\newblock Fourth-order finite-difference {P-SV} seismograms.
\newblock {\em Geophysics}, 53(11):1425--1436.

\bibitem[Luo and Schuster, 1991]{Luo_1991_WETc}
Luo, Y. and Schuster, G.~T. (1991).
\newblock Wave-equation traveltime inversion.
\newblock {\em Geophysics}, 56(5):645--653.

\bibitem[McMechan, 1983]{McMechan_1983_MET}
McMechan, G. (1983).
\newblock Migration by extrapolation of time-dependent boundary values.
\newblock {\em Geophysical Prospecting}, 31(3):413--420.

\bibitem[M\'etivier and Brossier, 2016]{Metivier_2015_TOO}
M\'etivier, L. and Brossier, R. (2016).
\newblock The {SEISCOPE} optimization toolbox: A large-scale nonlinear
  optimization library based on reverse communication.
\newblock {\em Geophysics}, 81(2):F11--F25.

\bibitem[M\'etivier et~al., 2016]{Metivier_2016_OTI}
M\'etivier, L., Brossier, R., M\'erigot, Q., Oudet, E., and Virieux, J. (2016).
\newblock An optimal transport approach for seismic tomography: Application to
  {3D} full waveform inversion.
\newblock {\em Inverse Problems}, 32(11):115008.

\bibitem[Mich\'{e}a and Komatitsch, 2010]{Michea_2010_AFP}
Mich\'{e}a, D. and Komatitsch, D. (2010).
\newblock Accelerating a 3{D} finite-difference wave propagation code using
  {GPU} graphics cards.
\newblock {\em Geophysical Journal International}, 182(1):389--402.

\bibitem[Minkoff, 2002]{minkoff2002spatial}
Minkoff, S.~E. (2002).
\newblock {Spatial parallelism of a 3D finite difference velocity-stress
  elastic wave propagation code}.
\newblock {\em SIAM Journal on Scientific Computing}, 24(1):1--19.

\bibitem[Nihei and Li, 2007]{Nihei_2007_FRM}
Nihei, K.~T. and Li, X. (2007).
\newblock Frequency response modelling of seismic waves using finite difference
  time domain with phase sensitive detection ({TD-PSD}).
\newblock {\em Geophysical Journal International}, 169:1069--1078.

\bibitem[Nocedal, 1980]{Nocedal_1980_UQN}
Nocedal, J. (1980).
\newblock {Updating Quasi-{N}ewton Matrices With Limited Storage}.
\newblock {\em Mathematics of Computation}, 35(151):773--782.

\bibitem[Nocedal and Wright, 2006]{Nocedal_2006_NOO}
Nocedal, J. and Wright, S.~J. (2006).
\newblock {\em Numerical Optimization}.
\newblock Springer, 2nd edition.

\bibitem[Pratt, 1999]{Pratt_1999_SWIb}
Pratt, R.~G. (1999).
\newblock Seismic waveform inversion in the frequency domain, {Part I}: theory
  and verification in a physical scale model.
\newblock {\em Geophysics}, 64:888--901.

\bibitem[Seriani and Priolo, 1994]{Seriani_1994_NSI}
Seriani, G. and Priolo, E. (1994).
\newblock Spectral element method for acoustic wave simulation in heterogeneous
  media.
\newblock {\em Finite elements in analysis and design}, 16:337--348.

\bibitem[Sirgue et~al., 2007]{Sirgue_2007_SM3D}
Sirgue, L., Etgen, T.~J., Albertin, U., and Brandsberg-Dahl, S. (2007).
\newblock System and method for 3{D} frequency-domain waveform inversion based
  on 3{D} time-domain forward modeling.
\newblock {\em {US} Patent Application Publication}, {US}2007/0282535 A1.

\bibitem[Smith, 1975]{smith1975application}
Smith, W.~D. (1975).
\newblock The application of finite element analysis to body wave propagation
  problems.
\newblock {\em Geophysical Journal International}, 42(2):747--768.

\bibitem[Tarantola, 1984]{Tarantola_1984_ISR}
Tarantola, A. (1984).
\newblock Inversion of seismic reflection data in the acoustic approximation.
\newblock {\em Geophysics}, 49(8):1259--1266.

\bibitem[Virieux, 1984]{Virieux_1984_SWP}
Virieux, J. (1984).
\newblock {SH} wave propagation in heterogeneous media: {V}elocity-stress
  finite difference method.
\newblock {\em Geophysics}, 49:1259--1266.

\bibitem[Virieux, 1986]{Virieux_1986_WPH}
Virieux, J. (1986).
\newblock {P-SV} wave propagation in heterogeneous media: {V}elocity stress
  finite difference method.
\newblock {\em Geophysics}, 51:889--901.

\bibitem[Virieux et~al., 2011]{Virieux_2011_RSP}
Virieux, J., Calandra, H., and Plessix, R.~E. (2011).
\newblock A review of the spectral, pseudo-spectral, finite-difference and
  finite-element modelling techniques for geophysical imaging.
\newblock {\em Geophysical Prospecting}, 59:794--813.

\bibitem[Virieux and Operto, 2009]{Virieux_2009_OFW}
Virieux, J. and Operto, S. (2009).
\newblock An overview of full waveform inversion in exploration geophysics.
\newblock {\em Geophysics}, 74(6):WCC1--WCC26.

\bibitem[Warner and Guasch, 2016]{Warner_2016_AWI}
Warner, M. and Guasch, L. (2016).
\newblock Adaptive waveform inversion: Theory.
\newblock {\em Geophysics}, 81(6):R429--R445.

\bibitem[Weiss and Shragge, 2013]{weiss2013solving}
Weiss, R.~M. and Shragge, J. (2013).
\newblock Solving 3d anisotropic elastic wave equations on parallel gpu
  devices.
\newblock {\em Geophysics}, 78(2):F7--F15.

\bibitem[Yang, 2023a]{Yang_2023_3dcsem}
Yang, P. (2023a).
\newblock {3D fictitious wave domain CSEM inversion by adjoint source
  estimation}.
\newblock {\em Computers \& Geosciences}, 180:105441.

\bibitem[Yang, 2023b]{Yang_2023_libEMM}
Yang, P. (2023b).
\newblock {libEMM: A fictious wave domain 3D CSEM modelling library bridging
  sequential and parallel GPU implementation}.
\newblock {\em Computer Physics Communications}, 288:108745.

\bibitem[Yang, 2023c]{Yang_2024_formulation}
Yang, P. (2023c).
\newblock A unified formulation of nonlinear, linearized and reflection
  waveform inversion.
\newblock {\em Bulletin of the Seismological Society of America}, page under
  review.

\bibitem[Yang et~al., 2016a]{Yang_2016_CAR}
Yang, P., Brossier, R., M\'etivier, L., and Virieux, J. (2016a).
\newblock Wavefield reconstruction in attenuating media: A
  checkpointing-assisted reverse-forward simulation method.
\newblock {\em Geophysics}, 81(6):R349--R362.

\bibitem[Yang et~al., 2018]{Yang_2018_TRN}
Yang, P., Brossier, R., M\'etivier, L., Virieux, J., and Zhou, W. (2018).
\newblock A {T}ime-{D}omain {P}reconditioned {T}runcated {N}ewton {A}pproach to
  {V}isco-acoustic {M}ultiparameter {F}ull {W}aveform {I}nversion.
\newblock {\em SIAM Journal on Scientific Computing}, 40(4):B1101--B1130.

\bibitem[Yang et~al., 2016b]{Yang_2016_WRB}
Yang, P., Brossier, R., and Virieux, J. (2016b).
\newblock Wavefield reconstruction from significantly decimated boundaries.
\newblock {\em Geophysics}, 80(5):T197--T209.

\bibitem[Yang et~al., 2014]{Yang_2014_RTM}
Yang, P., Gao, J., and Wang, B. (2014).
\newblock {RTM} using effective boundary saving: A staggered grid {GPU}
  implementation.
\newblock {\em Computers \& Geosciences}, 68:64--72.

\bibitem[Yang et~al., 2015]{Yang_2015_GPU}
Yang, P., Gao, J., and Wang, B. (2015).
\newblock A graphics processing unit implementation of time-domain
  full-waveform inversion.
\newblock {\em Geophysics}, 80(3):F31--F39.

\bibitem[Zhang and Zhang, 2008]{Zhang_2008_RTM}
Zhang, H.~J. and Zhang, Y. (2008).
\newblock Reverse time migration in 3{D} heterogeneous {TTI} media.
\newblock {\em SEG Technical Program Expanded Abstracts}, 27(1):2196--2200.

\bibitem[Zhang and Sun, 2009]{Zhang_2009_PIR}
Zhang, Y. and Sun, J. (2009).
\newblock Practical issues in reverse time migration: true amplitude gathers,
  noise removal and harmonic source encoding.
\newblock {\em First break}, 27:53--59.

\bibitem[Zhang et~al., 2023]{zhang2023nature}
Zhang, Z., Irving, J., Simons, F., and Alkhalifah, T. (2023).
\newblock Seismic evidence for a 1000 km mantle discontinuity under the
  pacific.
\newblock {\em Nature Communications}, 14.

\end{thebibliography}
\newcommand{\SortNoop}[1]{}

\end{document}